\documentclass[acmsmall,screen]{acmart}
\usepackage{microtype}

\setcopyright{rightsretained}
\acmJournal{PACMPL}
\acmYear{2020} \acmVolume{4} \acmNumber{OOPSLA} \acmArticle{189} \acmMonth{11} \acmPrice{}\acmDOI{10.1145/3428257}

\usepackage{booktabs}   %% For formal tables:
\usepackage{subcaption} %% For complex figures with subfigures/subcaptions
\usepackage{graphicx}
\usepackage{program}
\usepackage{centernot}
\usepackage{xcolor}
\usepackage[normalem]{ulem}
\usepackage{xifthen}
\usepackage{xparse}
\definecolor{codegreen}{rgb}{0,0.6,0}
\definecolor{codegray}{rgb}{0.5,0.5,0.5}
\definecolor{codepurple}{rgb}{0.58,0,0.82}
\definecolor{backcolour}{rgb}{1.0,1.0,1.0}
\usepackage{realboxes}
\usepackage{listings}
\usepackage{wrapfig}
\lstdefinestyle{pstyle}{
    backgroundcolor=\color{backcolour},   
    commentstyle=\color{codegreen},
    keywordstyle=\color{magenta},
    numberstyle=\tiny\color{codegray},
    stringstyle=\color{codepurple},
    basicstyle=\ttfamily\scriptsize,
    breakatwhitespace=false,         
    breaklines=true,                 
    captionpos=b,                    
    keepspaces=true,                 
    numbers=left,                    
    numbersep=5pt,                  
    showspaces=false,                
    showstringspaces=false,
    showtabs=false,                  
    tabsize=2,
	mathescape=true,
	escapechar=|
}

\lstdefinestyle{pnnstyle}{
    backgroundcolor=\color{backcolour},   
    commentstyle=\color{codegreen},
    keywordstyle=\color{magenta},
    stringstyle=\color{codepurple},
    basicstyle=\ttfamily\scriptsize,
    breakatwhitespace=false,
	numbers=none,    
    breaklines=true,                 
    captionpos=b,                    
    keepspaces=true,                                
    showspaces=false,                
    showstringspaces=false,
    showtabs=false,                  
    tabsize=2,
	mathescape=true,
	escapechar=|
}

\newenvironment{itemize*}%
  {\begin{itemize}%
}
  {\end{itemize}}
\newenvironment{enumerate*}%
  {\begin{enumerate}%
}
  {\end{enumerate}}
 
\newcommand\textts[1]{\texttt{\small #1}}
  \newcommand\ignore[1]{}
\newcommand\red[1]{\textcolor{red}{#1}}

\DeclareDocumentCommand{\diff}{o m}{#2}

\newcommand{\tool}{\textsf{DynamiTe}}
\newcommand{\UAutomizer}{\textsf{UAutomizer}}
\newcommand\eg{{\it e.g.}}
\newcommand\ie{{\it i.e.}}
\newcommand\etal{{\it et al.}}

\newcommand{\chanh}[1]{\noindent{\color{cyan} Chanh: #1}}

\newcommand\aut{\mathcal{A}}
\newcommand\sem[1]{[\![ #1 ]\!]}
\newcommand\run{r}
\newcommand\Cond{\mathcal{C}}
\newcommand\Rel{\mathcal{T}}

\newcommand\cfarightarrow[1]{\xrightarrow{#1}}
\newcommand\vln{\theta}
\newcommand\Vlns{\Theta}

\newcommand\ttskip{\texttt{skip}}
\newcommand\ttassume[1]{\texttt{assume}(#1)}

\newcommand\trace{\pi}
\newcommand\traces{\bar{\trace}}
\newcommand\traceBase{\trace_{\texttt{base}}}
\newcommand\traceTerm{\trace_{\texttt{term}}}
\newcommand\traceLoop{\trace_{\texttt{mayloop}}}
\newcommand\tracesBase{\traces_{\texttt{base}}}
\newcommand\tracesTerm{\traces_{\texttt{term}}}
\newcommand\tracesLoop{\traces_{\texttt{mayloop}}}
\newcommand\card[1]{\mathbf{card}(#1)}

\newcommand\GenInput{\texttt{GuessInput}}

\newcommand\DInfer{\texttt{DynInfer}}
\newcommand\Refine{\RefineRS}
\newcommand\ProveT{\textsf{ProveT}}
\newcommand\ProveNT{\textsf{ProveNT}}
\newcommand\ProveTNT{\textsf{ProveTNT}}
\newcommand\Pinstr{P_{\texttt{instr}}}
\newcommand\InferRF{\textsf{InferRF}}
\newcommand\IsUnchanged{\texttt{IsUnchanged}}
\newcommand\ValidateRFs{\texttt{ValidateRFs}}
\newcommand\GuessInputs{\texttt{GuessInputs}}
\newcommand\Execute{\texttt{Execute}}
\newcommand\Project{\texttt{Project}}
\newcommand\Partition{\texttt{Partition}}
\newcommand\GenTCTrans{\texttt{GenTCTrans}}
\newcommand\IsEmpty{\texttt{IsEmpty}}
\newcommand\RandPop{\texttt{RandPop}}
\newcommand\GenRFTemplate{\texttt{GenRFTemplate}}
\newcommand\Solve{\texttt{Solve}}
\newcommand\NotSatisfied{\texttt{NotSatisfied}}
\newcommand\RefineRS{\texttt{RefineRS}}
\newcommand\IsSat{\texttt{IsSat}}
\newcommand\IsValid{\texttt{IsValid}}
\newcommand\Term{\texttt{Term}}
\newcommand\GetLoopSeq{\texttt{GetLoopSeq}}
\newcommand\NonTerm{\texttt{NonTerm}}
\newcommand\Unk{\texttt{Unk}}

\newcommand\vtracet{{\bf vtrace}}

\newcommand\vtracePre[1]{\vtracet^{#1}_{\texttt{pre}}}
\newcommand\vtraceBody[1]{\vtracet^{#1}_{\texttt{body}}}
\newcommand\vtracePost[1]{\vtracet^{#1}_{\texttt{post}}}
\newcommand\vtracePreS{\vtracet_{\texttt{pre}}}
\newcommand\vtraceBodyS{\vtracet_{\texttt{body}}}
\newcommand\vtracePostS{\vtracet_{\texttt{post}}}

\newcommand\true{\textsf{true}}

\newcommand\Tstem{\Rel_{\texttt{stem}}}
\newcommand\Cloop{\Cond_{\texttt{loop}}}
\newcommand\Tloop{\Rel_{\texttt{loop}}}
\newcommand\Vars{\bar{V}}
\newcommand\Cterm{\Cond_{\texttt{term}}}
\newcommand\Cmayloop{\Cond_{\texttt{mayloop}}}

\newcommand\materials{supplemental materials~\cite{github}}

\begin{document}

%% Title information
%\title[Short Title]{{\sc Dynamite}:  Drawing Dynamic Termination and Non-termination Proofs Together}
%\title[\tool]{\tool: From Counterexamples to Sampling for Dynamic Termination and Non-termination Proofs}
%\title[\tool]{\tool: Crossing Both Lines for (Non)Termination Proofs}
%\title[\tool]{\resizebox{\textwidth}{!}{\tool: Dynamic Termination and Non-Termination Proofs}}
\title{\tool: Dynamic Termination and Non-Termination Proofs}
        %% [Short Title] is optional;
                                        %% when present, will be used in
                                        %% header instead of Full Title.
%\titlenote{with title note}             %% \titlenote is optional;
                                        %% can be repeated if necessary;
                                        %% contents suppressed with 'anonymous'
%\subtitle{Subtitle}                     %% \subtitle is optional
%\subtitlenote{with subtitle note}       %% \subtitlenote is optional;
                                        %% can be repeated if necessary;
                                        %% contents suppressed with 'anonymous'

%% Author information
%% Contents and number of authors suppressed with 'anonymous'.
%% Each author should be introduced by \author, followed by
%% \authornote (optional), \orcid (optional), \affiliation, and
%% \email.
%% An author may have multiple affiliations and/or emails; repeat the
%% appropriate command.
%% Many elements are not rendered, but should be provided for metadata
%% extraction tools.

%% Author with single affiliation.
\author{Ton Chanh Le}
%\authornote{with author1 note}          %% \authornote is optional;
                                        %% can be repeated if necessary
%\orcid{nnnn-nnnn-nnnn-nnnn}             %% \orcid is optional
\affiliation{
%  \position{Position1}
%  \department{Department1}              %% \department is recommended
  \institution{Stevens Institute of Technology}            %% \institution is required
%  \streetaddress{Street1 Address1}
%  \city{City1}
%  \state{State1}
%  \postcode{Post-Code1}
  \country{USA}                    %% \country is recommended
}
\email{letonchanh@gmail.com}          %% \email is recommended

\author{Timos Antonopoulos}
%\authornote{with author1 note}          %% \authornote is optional;
%% can be repeated if necessary
%\orcid{nnnn-nnnn-nnnn-nnnn}             %% \orcid is optional
\affiliation{
  %  \position{Position1}
  %  \department{Department1}              %% \department is recommended
  \institution{Yale University}            %% \institution is required
  %  \streetaddress{Street1 Address1}
  %  \city{City1}
  %  \state{State1}
  %  \postcode{Post-Code1}
  \country{USA}                    %% \country is recommended
}
\email{timos.antonopoulos@yale.edu}          %% \email is recommended

\author{Parisa Fathololumi}
%\authornote{with author1 note}          %% \authornote is optional;
%% can be repeated if necessary
%\orcid{nnnn-nnnn-nnnn-nnnn}             %% \orcid is optional
\affiliation{
  %  \position{Position1}
  %  \department{Department1}              %% \department is recommended
  \institution{Stevens Institute of Technology}            %% \institution is required
  %  \streetaddress{Street1 Address1}
  %  \city{City1}
  %  \state{State1}
  %  \postcode{Post-Code1}
  \country{USA}                    %% \country is recommended
}
\email{pfatholo@stevens.edu}          %% \email is recommended

\author{Eric Koskinen}
%\authornote{with author1 note}          %% \authornote is optional;
%% can be repeated if necessary
%\orcid{nnnn-nnnn-nnnn-nnnn}             %% \orcid is optional
\affiliation{
  %  \position{Position1}
  %  \department{Department1}              %% \department is recommended
  \institution{Stevens Institute of Technology}            %% \institution is required
  %  \streetaddress{Street1 Address1}
  %  \city{City1}
  %  \state{State1}
  %  \postcode{Post-Code1}
  \country{USA}                    %% \country is recommended
}
\email{eric.koskinen@stevens.edu}          %% \email is recommended

\author{ThanhVu Nguyen}
%\authornote{with author1 note}          %% \authornote is optional;
%% can be repeated if necessary
%\orcid{nnnn-nnnn-nnnn-nnnn}             %% \orcid is optional
\affiliation{
  %  \position{Position1}
  %  \department{Department1}              %% \department is recommended
  \institution{University of Nebraska, Lincoln}            %% \institution is required
  %  \streetaddress{Street1 Address1}
  %  \city{City1}
  %  \state{State1}
  %  \postcode{Post-Code1}
  \country{USA}                    %% \country is recommended
}
\email{tnguyen@cse.unl.edu}          %% \email is recommended

%% Author with two affiliations and emails.
\renewcommand{\shortauthors}{T. C. Le, T. Antonopoulos, P. Fathololumi, E. Koskinen, T. Nguyen}

%% Abstract
%% Note: \begin{abstract}...\end{abstract} environment must come
%% before \maketitle command
\begin{abstract}
%!TEX root = ./paper.tex

%Recent years have seen a variety of (mostly static) techniques that look beyond linear arithmetic domains and aim to reason about the safety or termination/non-termination of programs with non-linear behavior.  Meanwhile, techniques are also increasingly exploiting dynamic analysis to bolster verification and recent algorithms have been able to guess invariants for programs that have non-linear behavior. 
%
There is growing interest in termination reasoning for non-linear programs and,
meanwhile, recent dynamic strategies have shown they are able to infer invariants
for such challenging programs.
These advances led us to hypothesize that perhaps such dynamic strategies for non-linear invariants could be adapted to learn recurrent sets (for non-termination) and/or ranking functions (for termination).

In this paper, we exploit dynamic analysis and draw termination and non-termination as well as static and dynamic strategies closer together in order to tackle non-linear programs. 
%First we give novel algorithms for proving them individually, each based on categorizing terminating versus potentially non-terminating executions.
%
For termination, our algorithm infers ranking functions from concrete transitive closures, and, for non-termination, the algorithm iteratively collects executions and dynamically learns conditions to refine recurrent sets.
%\ProveT\ 
%samples state pairs from the concrete transitive closure of loop bodies, and fitting those pairs to a ranking function template, using SMT provers to solve for unknown coefficients. The resulting candidate ranking functions may be validated or else, if they are invalid, counterexamples are used to dynamically sample more executions for refinement.
%%
%Meanwhile, \ProveNT\ iconditions for terminating-versus-nonterminating to refine candidate recurrent sets. 
%
Finally, we describe an integrated algorithm that allows these algorithms to mutually inform each other, taking counterexamples from a failed validation in one endeavor and crossing both the static/dynamic and termination/non-termination lines, to create new execution samples for the other one.

We have implemented these algorithms in a new tool called \tool.  For non-linear programs, there are currently no  SV-COMP termination benchmarks so we created new sets of 38 terminating and 39 non-terminating programs.
Our empirical evaluation shows that we can effectively guess (and sometimes even validate) ranking functions and recurrent sets for programs with non-linear behaviors. Furthermore, we show that counterexamples from one failed validation can be used to generate executions for a dynamic analysis of the opposite property.
Although we are focused on non-linear programs, as a point of comparison, we compare \tool's performance on linear programs with that of the state-of-the-art tool, Ultimate.
%we applied our tool SVCOMP \texttt{termination-crafted-lit}, and compared against a state-of-the-art termination prover: Ultimate. 
Although \tool\ is an order of magnitude slower it is nonetheless somewhat competitive and sometimes finds ranking functions where Ultimate was unable to. Ultimate cannot, however, handle the non-linear programs in our new benchmark suite.

{\bf Supplemental Materials}. Our materials are available at
%\url{https://anonymous.4open.science/repository/af74f045-fcc5-485d-a56e-924eae39b3cc/README.md}, using the anonymizing GitHub web site 4open\footnote{Since these materials include source code and benchmarks, we have made them available on the anonymizing web site \url{https://anonymous.4open.science/}. We can see the number of views/visitors, but are not told where the traffic comes from. You may also wish to use a browser privacy mode.}.
\url{https://github.com/letonchanh/dynamite}.

%to 66 \emph{linear} programs taken from the SV-COMP benchmark \texttt{termination-crafted-lit}, and compared against a state-of-the-art termination prover: Ultimate. 

\end{abstract}

%% 2012 ACM Computing Classification System (CSS) concepts
%% Generate at 'http://dl.acm.org/ccs/ccs.cfm'.

%% End of generated code

%% Keywords
%% comma separated list
%\keywords{keyword1, keyword2, keyword3}  %% \keywords are mandatory in final camera-ready submission

%% \maketitle
%% Note: \maketitle command must come after title commands, author
%% commands, abstract environment, Computing Classification System
%% environment and commands, and keywords command.

\maketitle

%!TEX root = ./paper.tex

\section{Introduction}

Termination continues to be an important theoretical property that is
of practical interest. 
In recent years, there has been a proliferation of termination \diff{and non-termination}
verification tools, including
{\sc T2}~\cite{T2},
Ultimate~\cite{ultimate},
CPAchecker~\cite{beyer2011cpachecker},
AProVE~\cite{giesl2014proving},
FuncTion~\cite{urban2015function},
{\sc SeaHorn}~\cite{seahorn},
\diff{{\sc HipTNT+}}~\cite{Le2015},
and many others
(see the Termination track of SV-COMP~\cite{svcomp2020}). 
These tools are very effective at \diff{proving} termination \diff{and non-termination}, especially for
programs with linear arithmetic assignments and loop guards~\cite{Podelski2004}.

Meanwhile, researchers are increasingly using techniques based on dynamic execution, to bolster static verification. 
Static analysis explores all possible program paths but typically has one or more  shortcomings:  expressivity sacrifices, false positives, simpler invariants or restrictions on kinds of target programs. 
%False positive\timos{to what question? is this only when over-approximating?} rates are also problematic, making static techniques burdensome to users.  
Dynamic analysis focuses on exploring only a few program executions and, as such, \diff{also} has its own shortcomings: it is only correct with respect to the explored paths.  However, by (initially) sacrificing soundness, dynamic analyses support more expressive invariants and scale well to large and complex programs (see, \eg,~\cite{DBLP:journals/pacmpl/OHearn20}), often being  effective even when the source code is not available.  Moreover, false positives for the existence of a bug are not a concern: if any of the explored paths leads to an error, then it is a real error. 
More recent dynamic analyses have taken a ``data-driven'' or machine learning approach, \ie, learning based on training data.
%Recently, we also see many new approaches using dynamic analyses  in the form of ``data-driven'' or Machine Learning (i.e., they learn things based on the quality of train data and have limited soundness argument).\timos{Previous sentence has to be rephrased}
DIG~\cite{tosem2013} and ~\cite{yao2020learning} use this form of dynamic analysis to learn invariants of non-linear programs.
Moreover, other recent works combine both dynamic and static analysis techniques in an iterative loop, sometimes for the purpose of termination reasoning~\cite{Nori2013}. The dynamic analysis component is used to ``guess'' some candidate results and the static analysis one is used to verify them. The results of the checker, \eg, counterexamples showing invalidity of the candidate results, are then used to help the dynamic analysis to infer better results.

%This alternation between dynamic and static analysis techniques continues until a proof is found or a decision is made to stop with partial information.

%%%%%%%%%%%%%%%%%%%%%%%%%%%%%%%%%%%%%%%%%%%%%%%%%%%%%%%%
\emph{Landscape.} In the context of these recent advances that use dynamic support to learn invariants of non-linear programs, a natural question is whether they can be used or adapted to empower termination and non-termination reasoning for such challenging programs.
We explored in this direction, asking first whether non-termination reasoning can be built from dynamic approaches for non-linear invariants, and then a similar question for termination.
% Similarly, we then explored termination and, finally, we asked whether feedback from one endeavor could 
While these endeavors at first seem independent and could potentially be parallelized, we finally explored smarter approaches, where counterexamples from a failed termination proof could be used to generate executions for dynamically learning non-termination, and vice-versa.

%%%%%%%%%%%%%%%%%%%%%%%%%%%%%%%%%%%%%%%%%%%%%%%%%%%%%%%%
\emph{Learning \diff[rank]{ranking} functions and recurrent sets.}
In this paper, we present  algorithms that mix static and dynamic strategies in order to prove termination and non-termination of non-linear programs. Overall, our strategy begins by sampling terminating and potentially non-terminating executions from an instrumented program, with truncated divergence.
We first present a novel termination algorithm \ProveT, which dynamically samples \diff[pairs of]{} concrete states from the transitive closure of the loop bodies of sampled traces, fits them to a ranking function template, \diff[using]{and uses} an SMT solver to generate unknown coefficients in the template. We then attempt to validate these  candidate ranking functions (via reachability) and use any possible counterexamples to dynamically generate more sample traces. 
A second \ProveNT\ algorithm is used for non-termination, and iteratively refines a recurrent set condition, by executing the program and using samples to learn conditions for re-entering versus exiting the loop.
We next widen the interfaces of these procedures and show that these algorithms can inform each other. A failed attempt of \ProveT\ to prove termination yields a potentially infinite path, which we use to generate new executions to input to \ProveNT, and vice-versa.

%
%INTEGRATED
%dynamic can help T and NT separately.
%dynamic alone not validated.
%use static validation.
%sometimes that fails.
%use failed static validation of T to to create traces for dynamic NT
%use failed static validation of NT to to create traces for dynamic T

We have implemented these algorithms in a new tool called \tool\ for \diff{dynamically} proving termination and non-termination. \tool\ employs the power of many disparate tools:
the dynamic analysis tool DIG~\cite{tosem2013}, 
the \diff[static analysis]{symbolic execution} tool CIVL~\cite{civl},
the reachability analysis tools CPAchecker~\cite{beyer2011cpachecker} and Ultimate~\cite{ultimate} (without termination reasoning),
and the SMT solvers Z3~\cite{z3} and CVC4~\cite{cvc4}.

Our main goal is to prove termination and non-termination of non-linear programs, a domain for which many existing tools struggle.
\diff[Currently, there are no benchmark challenge programs for this domain. Our next contribution is to create such benchmarks]{
We evaluate {\tool} on two existing benchmarks consisting of non-linear programs (\textts{polyrank}~\cite{Bradley2005poly}, which has non-deterministic terminating programs, and {\sc Anant}~\cite{Cook2014}, which has non-linear non-terminating programs) and also create and evaluate {\tool} on more challenging non-linear terminating and non-terminating benchmarks (by adapting the~\citet{nladigbench} consisting of programs having nonlinear polynomial invariants)}.
%\diff[Currently, there are no benchmark challenge programs for this domain. Our next contribution is to create such benchmarks]{In addition to the two existing benchmarks: \textts{polyrank} of nondeterministic terminating programs}\footnote{There is only one non-linear program in this benchmark.}~\cite{Bradley2005poly} \diff{and {\sc Anant} of nonlinear non-terminating program}~\cite{Cook2014}\diff{, we create new challenging benchmarks for this domain}, by adapting the  SV-COMP \textsf{nla-digbench} suite for non-linear \diff{polynomial} invariants.
We  show that \tool\ is able to learn rich ranking functions and recurrent sets for these programs that cannot be handled by tools like Ultimate. We also show that the integrated algorithm can \diff[sometimes improve performance]{choose the right algorithm, \ProveT\ or \ProveNT\, or use the counterexample from a failed proof to assist the other}. 

Although non-linear programs were our focus, we also compared our algorithms on \diff{linear programs} against a state-of-the-art termination prover: Ultimate~\cite{ultimate}.  We used the 62 benchmarks from the category \textts{termination-crafted-lit} in SV-COMP 2020 and show that, although \tool\ is typically an order of magnitude slower (owing to the need for \diff{program} execution), it is nonetheless competitive with Ultimate, which is a much more mature tool. Also, in some cases \tool\ is faster than Ultimate and in other cases \tool\ is able to learn ranking functions, where Ultimate is unable to do so.
%%
%%Ideas:\\
%%1. use numerical invariants to represent recurrent set , ranking function (?)
%%2. Use existing dynamic invariant generation technique to infer very expressive invariants (e.g., nonlinear).  Many of such invariants are required to prove terminations (and we have a set of benchmarks to show)
%%3. Use an existing verifier to prove candidate invariants,  and use the results to help invariants.  While existing termination tools have hard time dealing with these programs (because they need to infer a nonlinear ranking function),  existing static verifier tool (e.g., Ultimate) can easily check and prove these invariants, provided that they are given.  In other words, synthesizing invariants is harder than checking for them. {\tool} exploits this facts (and we use example to show that's the case,  Ultimate termination tool (Automizer?) fails to prove termination , return Unknown,  but with the help of {\tool} in finding nonlinear invariants, Ultimate verification tool (Taipan?)  can easily prove these nonlinear invariants and therefore prove termination.
%%At the end, return a proved results.  
%%Some additional technical details here...   

\paragraph{Contributions} In summary, we present:
\begin{enumerate}
\item A novel termination algorithm, based on sampling concrete states from the transitive closure and fitting to ranking function templates with SMT. (Section~\ref{sec:alg_t})
\item A novel non-termination algorithm, based on refining recurrent sets with conditions learned from dynamic executions of the program (Section~\ref{sec:alg_nt}).
\item An integrated algorithm for termination and non-termination, that uses counterexamples from one failed static validation attempt to generate executions for dynamic analysis of the other. (Section~\ref{sec:tnt})
\item A new publicly available tool called \citet{github} for termination/non-termination of non-linear programs. (Section~\ref{sec:impl})
\item Two new benchmark suites for SV-COMP: one for termination of non-linear programs, and one for non-termination of non-linear programs. (Section~\ref{sec:eval})
\item An experimental evaluation, demonstrating that \tool\ is able to learn and sometimes validate ranking functions and recurrent sets for non-linear programs. (Section~\ref{sec:eval})

\end{enumerate}

\paragraph{Related work.}
\diff[In Section~\ref{sec:related} we discuss related works. Many works fall into various intersections between three concerns. ... While the above works share some similarities with ours, they differ for one reason or another.]{
Similarly to {\tool}, several existing works support programs with non-linear properties.
\citet{Nori2013} show program termination by dynamically inferring (nonlinear, disjunctive) loop bounds from program execution traces.
\citet{Bradley2005poly,Bradley05vmcai} use finite difference trees to statically infer lexicographic polynomial ranking functions to prove termination of non-linear programs.
For non-termination analysis, ~\citet{Cook2014} uses abstract interpretation to over-approximate the non-linear programs and infer linear recurrent sets to prove program termination.  
\citet{Frohn2019} uses recurrence solvers to generate loop-free transitions so that paths to non-terminating loops can be discovered.
In contrast, {\tool} integrates existing dynamic invariant generation and static verification for dynamically analyze both program termination and non-termination from their concrete snapshots, and it can analyze many other programs that are not by these works (details in Sections~\ref{sec:eval} and~\ref{sec:related}).
Also, {\tool} can analyze termination properties for non-deterministic programs (similarly to~\cite{Nori2013,Bradley2005poly,Bradley05vmcai}, but it currently cannot handle non-termination for non-deterministic programs (see Sections~\ref{sec:background}, ~\ref{sec:alg_nt} and~\ref{subsec:nlant} for additional discussion and evaluation).
In Section~\ref{sec:related} we discuss these works and other general termination and non-termination techniques in more details.}

\newcommand\ultimate{{\sc Ultimate}}
\newcommand\seahorn{{\sc SeaHorn}}
\newcommand\aprove{AProVE}

\section{Overview}\label{sec:overview}

\newcommand\tta{\textts{a}}
\newcommand\tts{\textts{s}}
\newcommand\ttt{\textts{t}}
\newcommand\ttn{\textts{n}}

\begin{wrapfigure}{r}{0.40\columnwidth}
  \vspace{-0.8cm}
  \begin{lstlisting}[language=Python,style=pnnstyle]
    
    int a = 0, n = *;
    while ((a+1) * (a+1) <= n):
      a = a + 1
   \end{lstlisting}
   \vspace{-0.8cm}
 \end{wrapfigure}
Consider the program to the right.
In the loop body of this program, \tta\ is incremented by 1, and
 the loop terminates when the square of \tta\textts{+1} is no longer below \ttn.
While the termination of this program is intuitively obvious, existing tools
(\eg\ \ultimate, \aprove, \seahorn) are unable to prove it to be terminating because it
requires reasoning about the non-linear behavior of program variables. Proving termination here involves discovering a ranking function that pertains to variables occurring in a quadratic inequality in the loop condition. As we discuss below, examples like this and more complicated ones with polynomial expressions, foil many existing techniques that are based on linear arithmetic constraints. One major impediment appears to be the lack of static reasoning techniques for programs with such non-linear behaviors. 
Some static works have shown static termination reasoning for certain classes of non-linear programs (\eg\ ``NAW loops'' \cite{Babic2007}, \diff{loops with finite difference trees~\cite{Bradley2005poly,Bradley05vmcai})}, and other works provide static resource bounds~\cite{Hoffmann2010a,Hoffmann2010b,Hoffmann2011,gulwani2009speed, gulwani2009speedb},
but still lack general techniques for termination and non-termination of these challenging programs.

%The following similar program also evades existing tools:
%\begin{center}
%  \begin{program}[style=tt]
%    int s = 1, t = 1, n = *;
%    wh\tab ile ((t + 1)*(t + 1) <= 4*n) \{
%        t = t + 2;
%        s = s + t; \untab
%    \}
%\end{program}
%\end{center}
%Again here, proving termination requires reasoning about the fact that the square of \ttt\textts{+1} is increasing, but also the fact the bound is the product \textts{4*}\ttn.

Meanwhile, in recent years, a number of works have showed that dynamic analysis can be used to learn rich, non-linear \emph{invariants}. \citet{nguyen2012using} showed that we can use dynamic analysis to learn expressive (non-linear) polynomial invariants from a small set of program execution traces. Subsequent works~\cite{nguyen2017counterexample,nguyen2017syminfer} propose iterative loop algorithms to generate candidate invariants from traces and use symbolic execution to refute spurious results and generate valid counterexamples, which are then used to improve the invariant generation process.
Many other works, e.g.,~\cite{sharma2013data,nguyen2014using}, combine inferring non-linear invariants with static checking.
Recently, \citet{yao2020learning} proposes using neural networks to learn non-linear invariants.

\subsection{Learning ranking functions and recurrent sets}

In this paper, we explore the question of whether techniques for non-linear invariants can be extended to reasoning about both termination and non-termination and do so in an integrated way. We begin by adapting earlier dynamic analysis works, to provide a new route for learning ranking functions and recurrent sets. To highlight and illustrate the key features of our work, we will use the following pair of slightly more complicated examples. The following two programs are similar, but one terminates and the other does not:
\begin{center}
\begin{tabular}{|l|c|l|}
\cline{1-1} \cline{3-3}
{\bf Termination} & &  {\bf Non-Termination}\\
\cline{1-1} \cline{3-3}
\hspace{0.2in}\begin{minipage}{0.42\textwidth}
  \begin{lstlisting}[language=Python]
int s = 1, t = 1, k, c = 1
while (t*t - 4*s + 2*t + 1 + c <= k):
  t = t + 2
  s = s + t
  c = c + t\end{lstlisting}
\end{minipage}
&\hspace{0.14in}&
\hspace{0.2in}\begin{minipage}{0.42\textwidth}
% >>>>>>> 99075a21c66e3b3574a83cf06c4259eea45fe2a1
\begin{lstlisting}[language=Python]
int s = 1, t = 1, c = 1
while (t*t - 4*s + 2*t + 1 + c >= 0):
  t = t + 2
  s = s + t
  c = c + t
\end{lstlisting}
\end{minipage}\\
\cline{1-1} \cline{3-3}
\end{tabular}
\end{center}
\noindent
These examples are based on \textts{sqrt1.c} from the~\citet{nladigbench}. (We discuss how we adapted it in Section~\ref{sec:eval}.) Both of these programs involve loop conditions that are non-linear, given the quadratic term \textts{t*t} in them. For illustration purposes, the body of both programs is the same, and it is not difficult to see by induction that the subexpression \textts{t*t - 4*s + 2*t + 1} in both loop conditions is always equal to 0. 

In the program on the left, the loop condition is essentially equivalent to \textts{c <= k} for all reachable states in the program. Given that \textts{k} is initialized to a nondeterministic value and unchanged and \textts{t} is always positive, and thus \textts{c} is always increasing up to \textts{k} and the execution will eventually exit the loop. This reasoning is usually captured with ranking functions: a map from every state to an ordered element where a transition in the program between states implies a transition from an element to a strictly smaller in that order element. Moreover, such an order is chosen to not be forever decreasing, and thus there cannot be an infinite sequence of states with valid program transitions between them. In this case such a function would be the one that maps every state to the value of \textts{k-c}. It is the aim of our algorithm to synthesize this function and we describe in the following paragraphs how this is achieved.

In the program on the right, the loop condition is now essentially \diff[\textts{0 <= 0}]{\textts{c >= 0}} for all reachable states of the program, which holds trivially \diff{since \textts{c} is initialized to \textts{1} and always increasing}. To show non-termination, a recurrent set is usually constructed: by abstracting the loop body in a relation $\Tloop$, a set $X$ of states is collected such that for any state $s$ in that set, and any other state $s'$ we can transition to from $s$, it is the case that $s'$ is also in $X$. The existence of such a set that contains reachable states implies that the program is non-terminating at least in some cases. 

The above examples reiterate the point that, even for simple nonlinear programs, the necessary reasoning evades existing termination and non-termination tools, many of which are based on linear arithmetic constraint solving~\cite{Podelski2004,seahorn}.

%toward dynamic analysis.
%what's hard about this problem.
%It is the aim of our tool to construct such sets of states and we describe in what follows how this is performed.
%
%\red{first insight: use dynamic traces and inference to guess ranking functions. explain {\tool}}

% In the body of the loop, the variables are incremented.
% % T is incremented by two, while S is in a committed by T and see is incremented buy one.
% This loop is terminating: The expression \textts{t*t - 4*s + 2*t + 1}
% is equal to 0 on every iteration of the loop and the value of c monotonically increases, since \textts{t} is always positive.
% Therefore the whole left side of the inequality is monotonically increasing and is bounded by \textts{k}, which remains unchanged through out.

% Meanwhile, the program on the right is a small variant lasdkmfalsdkjfmlakdsm 

\paragraph{Dynamic snapshots for termination/non-termination.}
In this paper, we work in a direction that is based on learning from dynamically generated program traces, for guessing (and possibly validating) ranking functions and recurrent sets for examples such as those above.  To this end, we begin by describing a simple mechanism for collecting traces which may or may not terminate.
Tools for dynamic analysis typically take ``snapshots'' of the state of the program
to record trace information. These snapshots may record the values of some/all variables, as well as the program location and there are many techniques for injecting snapshots. 

In the case of termination, though, there is the additional challenge of recording snapshots of a loop that may run forever. Our solution is to break the loop so that potentially infinite executions are truncated, and then we can learn about those prefixes and try to characterize the ones that would have terminated versus those that would not have. We can truncate loops by introducing a counter. Counter instrumentation is common in other kinds of static analysis~\cite{gulwani2009speed, gulwani2009speedb,Hoffmann2010a,Hoffmann2010b,Hoffmann2011}, but here we use it to truncate executions.

\begin{wrapfigure}[10]{r}[34pt]{3.1in}
  \hspace{0.2in}
  \begin{minipage}{2.9in}
  \vspace{-0.2in}    
  \begin{lstlisting}[language=Python]
int s = 1, t = 1, c = 1
|\Colorbox{lightgray}{int \_ctr = 0, \_bnd = 500}|
|\Colorbox{lightgray}{$\vtracePreS$(s,t,c)}|
while (t*t - 4*s + 2*t + 1 + c >= 0):
  |\Colorbox{lightgray}{if (\_ctr > \_bnd) abort() else \_ctr++}|
  |\Colorbox{lightgray}{$\vtraceBodyS$(s,t,c)}|
  t = t + 2
  s = s + t
  c = c + t
|\Colorbox{lightgray}{$\vtracePostS$(s,t,c)}|
\end{lstlisting}
\end{minipage}
\end{wrapfigure}
\newcommand\ctr{\textts{\_ctr}}
\newcommand\bnd{\textts{\_bnd}}
Using the above non-termination example, our transformation generates the program on the right, where changes are indicated in the gray boxed regions.
%As illustrated in the boxed regions of the \diff[above]{} program \diff{on the right}, our transformation makes a few changes.
Technically, these changes are made to every loop of the program, but for illustrative purposes in the section we will just use one loop.
Our transformation begins by introducing a pair of variables for the loop. The variable \ctr\ is used to count the loop's iterations, and \bnd\ is an input to the dynamic analysis, whose value is pragmatically chosen to determine a useful prefix of potentially non-terminating traces. As such, \ctr\ is incremented inside the loop body and when it reaches \bnd, control exits the loop if it hasn't already.
Next, we inject function calls to a method $\vtracet$ in three places: before ($\vtracePreS$), during ($\vtraceBodyS$), and after ($\vtracePostS$) the loop. In each location, $\vtracet$ is used to record the values of the variables in scope as a tuple such as
$(\textts{body},\textts{s}=1,\textts{t}=1,\textts{c}=1)$, which we call a ``snapshot''. It should be noted that $\vtracePostS$ only captures values for states that exited the loop due to natural causes. 
%\eric{what do we do in the case where the loop would have exited from natural causes at exactly the same time that the counter reaches the bound?}

%\begin{center}
%  \begin{program}[style=tt]
%\boxedtt{int \_ctr = 0; int \_bound;
%\vtrace{0}(x, y);}
%wh\tab ile (x $>=$ 0) \{
%  \boxedtt{if (\_ctr $>$ \_bound) break;
%  else \_ctr++;
%  \vtrace{1}(x, y);}
%  x := x + y; \untab
%\}
%\boxedtt{if (\_ctr <= \_bound) \vtrace{2}(x, y);}
%  \end{program}
%\end{center}
%Alarmingly, this transformation {\bf changes the behavior of the program}.

\newcommand\pib{\tracesBase}
\newcommand\pit{\tracesTerm}
\newcommand\pil{\tracesLoop}
This truncation and instrumentation permits us to  distinguish between three classes of traces:
%\timos{Where did the $v_i$ come from?}\chanh{Should we instead use $s_i$ for ``snapshots""?}
%\[\begin{array}{ccc}
%\pib = \{ v0 \cdot v2 \}, &
%\pit = (v0 \cdot (v1)^+ \cdot v2), &
%\pil = (v0 \cdot (v1)^+)\\
%\end{array}\]
\[\arraycolsep=1.4pt\begin{array}{ccc}
\pib =  \{(\textts{pre},\_) \!\cdot\!  (\textts{post},\_)\}, &
\pit = \{(\textts{pre},\_) \!\cdot\!  (\textts{body},\_)^+ \!\cdot\!  (\textts{post},\_)\}, &
\pil = \{(\textts{pre},\_) \!\cdot\!  (\textts{body},\_)^+\}\\
\end{array}\]
Above we have described the classes of traces using simple regular expressions, matching the first component of the tuple, and ignoring the values of variables. Technically, by these expressions, we mean the set of all traces that match the regular expression.
The first set of traces $\pib$ are those that have a snapshot before the loop,  skip the loop entirely and then have a snapshot immediately after the loop. The second set $\pit$ is similar but has at least one snapshot from inside the body of the loop. The third set $\pil$ includes traces that entered the loop but for which there is no post-loop snapshot.
(Of course the union of these languages covers the language of the program.)
This transformation is unsound because (i) it does not account for loop body states beyond \bnd, and (ii) executions may be forced to exit the loop before their day has come to do so. However, as we will see, this strategy collects \emph{rich} data that enables us to start making guesses for ranking functions and recurrent sets, even in non-linear contexts such as this example.

\subsection{Algorithms}

\paragraph{Learning ranking functions for termination.}
In Section~\ref{sec:alg_t}, we present an algorithm beginning with:
\begin{enumerate*}
\item Instrumenting the program as discussed above.
\item Generating random inputs to the program  and collect traces (as in~\cite{nguyen2012using,tosem2013}).
\item Partitioning traces into $(\pib,\pit,\pil)$.
\item Using $\pit$ as an input to subprocedure \InferRF$(\pit)$, discussed below, to
infer a ranking function from the data.
\end{enumerate*}
For the above Termination example, such a possible trace is:
%(\textts{pre},1,1,1,1),
%(\textts{body},1,1,1,1),
%(\textts{body},4,3,1,4),
%(\textts{body},7,5,1,7),\ldots
%&\in&\pit\\
%\[\begin{array}{lll}
$\{(\textts{pre},1,1,42,1),
(\textts{body},1,1,42,1),$
$(\textts{body},4,3,42,4),
(\textts{body},9,5,42,9),\ldots \}
\in \pit$,
%\end{array}\]
where each tuple represents variables \textts{(\_,s,t,k,c)}.
Running \tool\ on this example takes 7.55 seconds to produce an answer. By comparison, existing tools for termination~\cite{ultimate,cpachecker,seahorn,aprove} typically perform well on linear programs, but fail to produce an answer on this program.
%\eric{run dynamo and see what RFs get emitted before validation.}
%While \eric{expores XXX traces, takes 7.55 seconds} This is significantly longer than it takes to verify linear arithmetic tools with existing tools such as~\cite{ultimate,cpachecker,seahorn,aprove}. Nonetheless those tools are unable to verify  \emph{non}-linear examples such as this one.
%
The output of \InferRF\ is the ranking function expression \textts{k-c}. In some cases this guess may already be useful, even though it has not been verified. A user may wish to examine it and, in this case, it appears to be correct. If a stronger guarantee is needed, this ranking function can be given to a \emph{safety} reachability prover such as (the reachability analyses of) CPAchecker~\cite{cpachecker} or Ultimate~\cite{ultimate}. For example, using a standard translation~\cite{Cook2006}, we can use an encoding that reduces ranking function validity checking to reachability. In this example, Ultimate's reachability reasoning can verify that \textts{k-c} is a valid ranking function after 167 seconds.

%%%analysis:936:DEBUG (prove_vloop) - Termination result (vloop_15): True ([k + -1*c])
%%%Termination result: True
%%%utils.profiling:31:DEBUG (timed) - prove: 175804.16ms
%%%Time log:
%%%gen_rand_inps: 0.174s
%%%_get_traces_mp: 0.242s
%%%_merge_traces: 0.191s
%%%get_traces_from_inps: 0.434s
%%%infer_ranking_functions: 7.552s
%%%prove_reach: 167.460s
%%%validate_ranking_functions: 167.616s
%%%prove: 175.804s

In some cases, however, \InferRF\ guesses incorrectly and returns a ranking function that is invalid. In these cases, a reachability solver may return a counterexample, which contains valuable information: a stem (path to the loop) and lasso (cycle through the loop body) that potentially non-terminates.
In Section~\ref{sec:alg_t} we describe a sub-procedure \GenInput(cex) that uses this counterexample to guide the generation of new program inputs that can lead to a trace corresponding to this stem and lasso.

% ./run svcomp-nla-digbench run dynamo sqrt1-both-t.c

%\paragraph{Guessing ranking functions (\InferRF)}
Our algorithm is parametric over the procedure \InferRF\ for inferring ranking functions from trace samples. In Section~\ref{sec:alg_t} we give one strategy that is based on taking samples from the transitive closure over the trace snapshots and then fitting them to a template, using models of the well-foundedness constraints from an SMT solver to generate unknown coefficients of the template. The output ranking functions can be seen in Section~\ref{sec:eval}.
%
% At the high level, for each trace, we randomly choose a sample of pairs of snapshot states for variables $\{v_1,\ldots,v_n\}$ with values $\{h_1,\ldots,h_n\}$ from the transitive closure of the concrete transition relation, in a manner analogous to disjunctive well-foundedness~\cite{djwf}.
% We next fit them to a template of the form\red{trim this. too much detail.}\timos{I agree}
% \[ u_0 + u_1 \cdot h_1^i + u_2 \cdot h^i_2 + \cdots + u_n \cdot h_n^i
% \]
% where $u_i$ is an unknown coefficient, and $h_i$ is the value of a corresonding.
% Using SMT, we look for a model that satisfies the constraints that
% (1) this expression is decreasing in this sample of the transitive closure
% and that (2) it is bounded from below by 0.
% \ie, this expression is a ranking function.

%if we get FALSE, we may then want to use the coutnerexample to start proving Non-termination (discussed below)

\emph{Learning recurrent sets for non-termination.}
%\label{sec:alg_nt}
For non-termination, the goal is to guess a \emph{recurrent set}, which is a set of states $X$, such that once $X$ is reached, every subsequent transition will return to $X$. (We will define this formally later.) We begin by instrumenting the program but, unlike the termination algorithm,  we instead  use a form of iterative refinement to learn recurrent sets. For each loop, our overall algorithm keeps a work-list of candidate recurrent sets, starting by using the loop \emph{condition} itself as \diff[a]{the first candidate} recurrent set. On each iteration, we select such a candidate recurrent set and check whether it is a valid recurrent set and reachable. If so, the algorithm has proven non-termination and returns. Otherwise, we have a model witness to the invalidity of the recurrent set which can be used for its refinement.

When $R$ is not a recurrent set, our procedure \Refine$(R,P,\Tstem,\Tloop)$ attempts to learn a refined set from traces of the program. \Refine\ uses the reason for the invalidity of $R$ to generate a set of traces $\Pi$ of the instrumented/truncated program. These traces are then used to learn conditions that lead to \emph{refined} candidate recurrent sets. These refined candidate recurrent sets are then returned to the outer algorithm for further validity checking.
%\red{show the rcr set generated. report times. valid?}

\emph{From failed validation to dynamic learning.}
%Integrating termination and non-termination.}
A failed proof of static termination can be used to inform a dynamic non-termination proof and vice-versa. We discuss how our algorithms can be integrated together in an algorithm called \ProveTNT, described in Section~\ref{sec:tnt}. The key idea is to additionally parameterize \ProveT\ (respectively, \ProveNT) by a set of input traces, which are derived from a failed \ProveNT\ (resp., \ProveT) attempt. These concrete traces contain useful data: examples of where the program appears to terminate or appears to diverge, and can immediately be used to guess ranking functions or recurrent sets.

\subsection{The \tool\ Tool}

We have developed \tool\ for dynamically guessing (and statically validating) rank
functions and recurrent sets.
The tool is publicly available at~\cite{github}.
%We plan to release \tool, and have included it in the
%Supplemental Materials~\cite{github}\tvn{plan to}.
%\begin{figure}[t]
\tool\ is written in a combination of Python and OCaml, the latter used mostly for program transformations (instrumentation and ranking function validity checking) with CIL~\cite{necula2002cil}.

\begin{wrapfigure}[16]{r}[34pt]{0.65\columnwidth}
  \includegraphics[width=0.55\columnwidth]{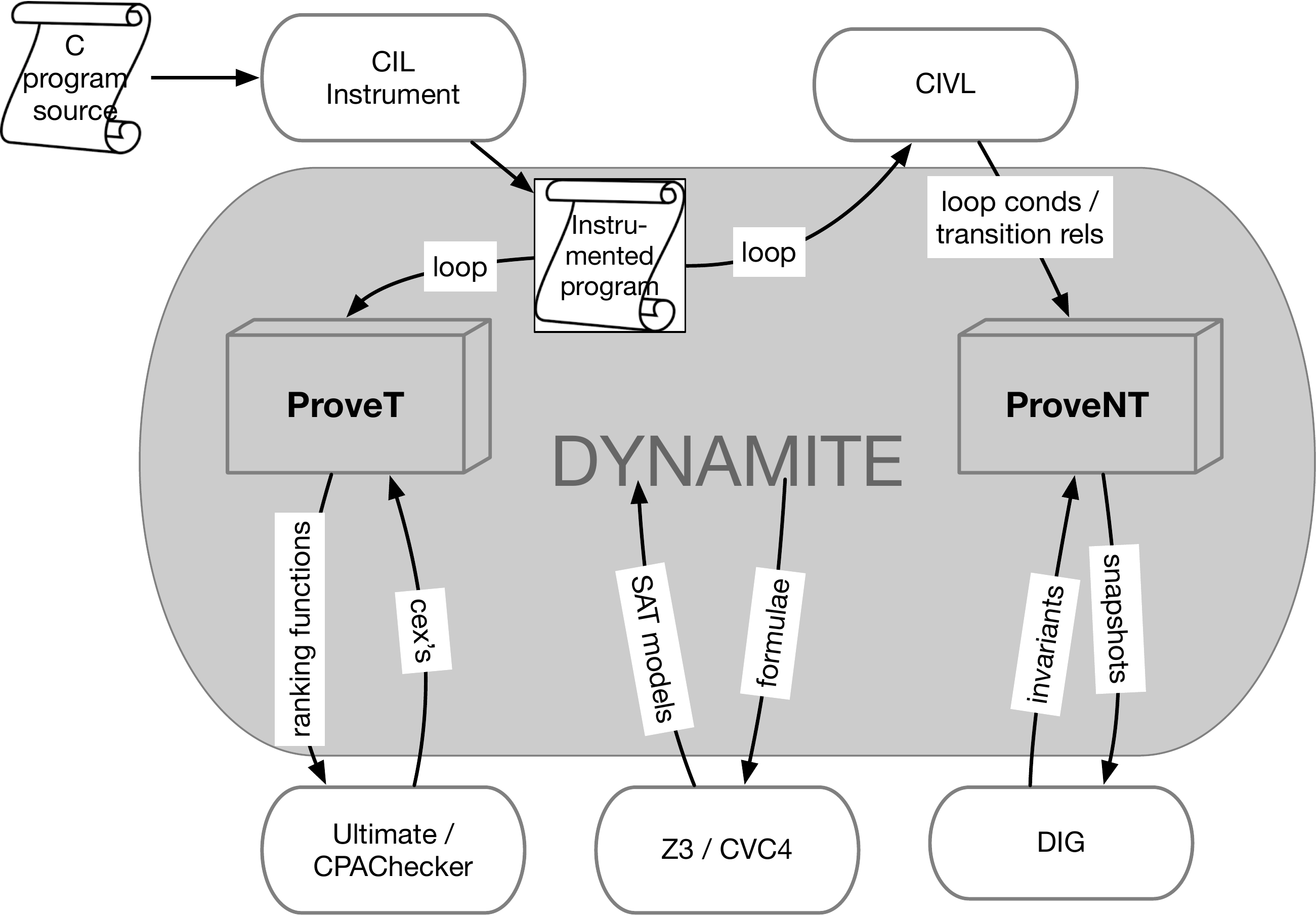}
  \caption{Tools used in {\tool}.}
  \label{fig:tools}
  %\end{figure}
\end{wrapfigure}
\tool\ takes advantage of several existing dynamic, symbolic, and static analysis techniques and tools as shown in Fig.~\ref{fig:tools}.
The two main algorithms to check for termination and nontermination mentioned above are the two blocks labeled \ProveT\ and \ProveNT, respectively.
As shown in the figure, \ProveT\ uses the two tools Ultimate and CPAChecker to verify the inferred ranking functions and obtain counterexamples to improve the inference process.
%\vu{chanh:  need more details: use Ultimate for what and CPAChecker for what?}
On the other hand, \ProveNT\ uses the CIVL symbolic execution tool to obtain program information such as loop conditions and transition relations, the DIG dynamic invariant generation tool to infer invariants from snapshot traces in order to represent and refine recurrent sets.
%\vu{everyone: check if what I say is true}.
\tool\ also uses the Z3 and CVC4 SMT solvers to check if the candidate recurrent sets are valid and if not obtains counterexamples to refine DIG's inference process.
%Thus, the integration and combination of external tools allow \tool\ to achieve what none of these tools can do individually.

%\paragraph{Sequential and nested loops.}
Our algorithms can work with programs containing \emph{sequential and/or nested loops}. Our program transformation puts each loop into a separate method and replaces the loop by a call to that method. We then build a call graph of those methods and extract a \emph{postorder} call sequence from it. We analyze each loop at a time in that order, i.e., the top-down innermost loop will be examined first. We proceed to the next loop in the sequence only when we have learned that the current loop is terminating. Otherwise, we conclude that the whole program does not terminate due to the non-termination of the current loop.

%% \tool\ uses the dynamic invariant tool DIG to learn nonlinear ranking invariants and the symbolic execution tool CIVL to obtain pathconditions\vu{Chanh, what is symexe used for in ProveNT?}.
%% To formally verify its results, \tool\ relies on static verifiers such as Ultimate Taipan and CPAChecker to prove learned ranking functions and Ultimate Automizer to formally prove termination results\vu{what about non-termination results?}.
%% \tool\ also uses Z3 and CVC4 for other miscelleneous operations such as implication checking and simplification.

Our main goal in this paper is to develop integrated termination/nontermination algorithms that exploit dynamic analysis to support non-linear programs. However, we also evaluate how \tool\ performs on \emph{linear} examples.
To this end, we evaluated \tool\ on the 66 benchmarks from the SV-COMP \textts{termination-crafted-lit} set of linear arithmetic termination and non-termination problems collected from literature. We compared \tool\ to Ultimate, because it is one of the most successful termination reasoning tools available. We report Ultimate's proving time as compared with \tool's guessing time and \tool's time to validate guesses. Details are in Section~\ref{sec:eval}. Overall, \tool\ is roughly an order of magnitude slower than Ultimate on linear benchmarks, owing to the fact that \tool\ must repeatedly execute the program to collect data. Nonetheless, it's worth noting that Ultimate is a much more mature tool. In one case, \tool\ was able to learn a rank function that Ultimate could not.  We also show that \tool\ is competitive for proving \emph{non}-termination of linear programs, as compared to Ultimate's ability to generate lasso counterexamples to termination.

For the non-linear case, there are \diff[no]{two} existing benchmarks: \diff[for termination/non-termination.]{\textts{polyrank} for termination and {\sc Anant} for non-termination. 
However, they only have at-most-quadratic polynomial programs and 10/11 programs in the \texttt{polyrank} benchmark are linear.} \diff[The closest relative is]{To make (non)termination reasoning more challenging, we adapted the closely related} SV-COMP \textts{digbench} set of programs for non-linear invariant generation problems \diff[We therefore adapted that set]{} to create two new sets of benchmarks called \textts{termination-nla} and \textts{nontermination-nla} which we are submitting to SV-COMP. 
The set \textts{termination-nla} consists of {37} terminating programs and \textts{nontermination-nla}  consists of {38} non-terminating programs, \diff[We adapted non-linear invariants in our creation of loop conditions, making (non)termination reasoning challenging]{which were created by adapting (up to sextic degree) non-linear invariants in their loop conditions}. 
Our empirical evaluation shows that \tool\ can discover and sometimes validate rank functions (in  35 of 37 cases) and recurrent sets (in 33 of 38 cases) for non-linear programs, that are not supported by Ultimate. (Ultimate returns an unsupported error message.)

%
%leads to the following conclusions:
%(i) \tool\ can handle linear\timos{what?} (maybe slower)
%(ii) \tool\ can handle some linear\timos{what?} that existing tools can't handle
%(iii) \tool\ can handle many nonlinear\timos{what?} that are out of reach of today's inference, but can still be validated
%(iv) \tool\ can guess nonlinear\timos{what?} that is even beyond today's validation.
%

%
%\red{relational instrumentation:}
%\begin{center}
%  \begin{program}[style=tt]
%\boxedtt{vtrace0(x, y);
%int counter = 0;}
%wh\tab ile (x $>=$ 0) \{
%  \boxedtt{if (counter $>$ bnd) break;
%  else counter++;
%  vtrace1(x, y);}
%
%  x0 = x;
%  y0 = y;
%  x1 = x0 + y0;
%  y1 = y0;
%  x = x1;
%  y = y1;
%  vtrace3(x0, y0, x1, y1);
%\}
%if\tab\ (counter <= bnd) 
%  vtrace2(x, y);
%  \end{program}
%\end{center}

%
%\paragraph{Example 2: Trigonometric functions.}
%We now describe a program for which proving termination/non-termination 
%escapes known techniques.
%
%Oftne programs, especially scientific applications, terminate due to more complicated mathematical reasons. Example:
%\begin{center}
%  \begin{program}[style=tt]
%    wh\tab ile (y < 42) \{
%      y := a * sin(x);
%      x := x + 1;
%      a := a + 1; \untab
%    \}
%  \end{program}
%\end{center}
%sine procedure. terminates when the amplitude goes above threshold of 42.
%
%On the other hand, classification based on dynamic analysis is very powerful and can infer the behavior
%
%

%!TEX root = ./paper.tex

\section{Preliminaries}
\label{sec:background}
%\begin{definition}[Transition system]
%A transition systems $M=(S,R,I)$ is a tuple consisting of a state space $S$, transition relation $R \subseteq S \times S$ and initial states $I \subseteq S$.
%\end{definition}

%\begin{definition}[Programs]
%We denote by $M_P$ the transition system induced by $P$.
%\end{definition}

We denote a program by $P$.  We assume, for simplicity, that it has a single set $V$ of variables. We will sometimes use the notation $V'$ to mean a second set of primed versions of the same variables, \ie\ $V' = \{ v' \mid v \in V\}$, to describe transition relations.
%set of variables $V$ and $V'$
We denote by $\Sigma$ set of states which we treat as valuations of the variables $V$, \ie\ $\Sigma: V\rightarrow Val$.
To represent conditions, we use logical formulae for states, denoted $\Cond$, where
$\sem{\Cond} : \Sigma \rightarrow \mathbb{B}$. We also work with logical
state \emph{transition relations} denoted $\Rel$, where $\sem{\Rel} \subseteq \Sigma \times \Sigma$. $\Rel$ can also be presented in the form of logical formulae. As we describe below, program loops can be summarized using these conditions and relations, in a standard way.

\begin{definition}[Ranking functions]
For a state space $S$, a ranking function $f$ is a total map 
from $S$ to a well-ordered set with ordering relation $\prec$. 
A relation $\Rel \subseteq S \times S$ is \emph{well-founded} if and only if there exists a ranking function $f$ such that 
$\forall (s,s') \in \Rel.\ f(s') \prec f(s)$.
\end{definition}

The existence of a ranking function over the transition relation $\Rel$ of a loop implies the termination of that loop, as there can be no infinite sequence of states $s_0,s_1,\ldots$ such that $\mathcal{T}(s_i,s_{i+1})$ holds for every $i\geq 0$. This is because for all $i\geq 0$, $f(s_{i+1})\prec f(s_i)$ and the sequence of states mapped under $f$ cannot be decreasing forever as the image of $f$ is a well-order.

The termination of a loop with a transition relation $\Rel$ can also be proved by a finite set of ranking functions (or measures) $\mathcal{M} = \{f_1, \ldots, f_m\}$ by showing that the transitive closure of $\Rel$ is contained in the \emph{disjunctively well-founded} relation defined from $\mathcal{M}$~\cite{Podelski2004a}. That is,
$ \Rel^+ \subseteq \{(s, s') ~|~ f_1(s') \prec f_1(s) \vee \ldots \vee f_m(s') \prec f_m(s)\} $. 

This validity of the finite set of ranking functions $\mathcal{M}$ for the loop's termination can be checked via proving safety of the following instrumented loop (i.e., the error is unreachable)~\cite{Cook2006}. This check can be performed by a reachability prover such as \citet{ultimate} or 
% \begin{wrapfigure}[10]{r}[34pt]{3.9in}
% \hspace{-0.2in}\begin{minipage}{4in}
%     \begin{lstlisting}[language=Python,style=pnnstyle]
%     |\boxedtt{\_dup = False}|
%     while C: |\fbox{\begin{minipage}{3.2in}
%       if \_dup:\\
%         if not ($f_1(\hat{x}_1, ..., \hat{x}_n) > f_1(x_1, ..., x_n)$ and $f_1(\hat{x}_1, ..., \hat{x}_n) \ge 0$):\\
%       $\ldots$\\
%       if not ($f_m(\hat{x}_1, ..., \hat{x}_n) > f_m(x_1, ..., x_n)$ and $f_m(\hat{x}_1, ..., \hat{x}_n) \ge 0$):\\
%       ERROR: skip\\
%       if not \_dup and *:\\
%       $\hat{x}_1 = x_1; ...; \hat{x}_n = x_n$\\
%       \_dup = True
%       \end{minipage}}|
%       B \end{lstlisting}
%   \end{minipage}
% \end{wrapfigure}
CPAchecker~\cite{beyer2011cpachecker}. 
Below is an illustration for a while loop, whose instrumentation code are put in gray boxes.

\begin{wrapfigure}[14]{r}[32pt]{4.0in}
  \hspace{-0.2in}\begin{minipage}{4in}
    \begin{lstlisting}[language=Python,style=pnnstyle]
      |\Colorbox{lightgray}{\_dup = False}|
      while C: 
        |\Colorbox{lightgray}{if \_dup:}|
          |\Colorbox{lightgray}{if not ($f_1(\hat{x}_1, ..., \hat{x}_n) > f_1(x_1, ..., x_n)$ and $f_1(\hat{x}_1, ..., \hat{x}_n) \ge 0$):}|
            |\Colorbox{lightgray}{$\ldots$}|
            |\Colorbox{lightgray}{if not ($f_m(\hat{x}_1, ..., \hat{x}_n) > f_m(x_1, ..., x_n)$ and $f_m(\hat{x}_1, ..., \hat{x}_n) \ge 0$):}|
              |\Colorbox{lightgray}{ERROR: skip}|
        |\Colorbox{lightgray}{if not \_dup and *:}|
          |\Colorbox{lightgray}{$\hat{x}_1 = x_1; ...; \hat{x}_n = x_n$}|
          |\Colorbox{lightgray}{\_dup = True}|
        B \end{lstlisting}
  \end{minipage}
\end{wrapfigure}
\noindent In this instrumented program, a state $\hat{s}$ of the loop is arbitrarily recorded and then for any subsequent state $s$, we check if the transition $(\hat{s}, s)$ satisfies at least one ranking function in $\mathcal{M}$. A transition $(\hat{s}, s)$ that does not satisfy any ranking function in $\mathcal{M}$, indicates that the transitive closure transition $\Rel^+$ is not a subset of the disjunctively well-founded relation of $\mathcal{M}$. In this case, the error is reached and the termination proof fails. Otherwise, a safe program in which the error is unreachable implies the loop's termination. 

\begin{definition}[Recurrent set]\label{def:rcr}
For sets of states $X$ and transition relation $\Rel$,
$X$ is a \emph{recurrent set} if
\[\begin{array}{lll}
\textrm{(1) } X\neq \emptyset, &
\textrm{(2) } \Rel \textrm{ is total on } X, &
\textrm{(3) } \textrm{ the image of } \Rel \textrm{ on } X \textrm{  is contained within } X
\end{array}\]
\end{definition}

%We will often work with recurrent sets and transition
%relations described with
%formulas (\eg\ denoted $\Cond$ and $\Rel$, respectively).

\diff{The above notion of recurrent sets (\ie\ ``closed recurrent sets'' in~\cite{Chen2014}) can help to avoid the difficulty and inefficiency of reasoning the $\forall\exists$ alternation in ``open recurrent sets''~\cite{Gupta2008}, but it cannot support non-determinism without under-approximation. Therefore, our non-termination proofs are restricted to only deterministic programs. Finding under-approximation of non-determinism from concrete possibly-nonterminating snapshots to support non-termination proofs of non-deterministic programs will be our future work. Note that such restriction does not apply to our termination proofs.}

\begin{definition}[Loop summary]
As is typical~\cite{Cook2006}, we will describe loops in terms
of a triple ($\Tstem,\Cloop,\Tloop$), where
$\Tstem$ over-approximates the transition from the entrypoint
of the program up to the loop header,
$\Cloop$ over-approximates the condition for entering the loop, and
$\Tloop$ over-approximates the transition through the entire body of the
loop back to the header.
\end{definition}

%!TEX root = ./paper.tex

\section{Inferring Ranking Functions for Termination}\label{sec:alg_t}

%\chanh{The loop $L$ is reachable since the termination of previous loops have been proved}

\begin{figure*}\centering
\begin{tabular}{ll}
	\hspace{0.02\textwidth}
\begin{minipage}{0.47\textwidth}
\begin{lstlisting}[language=Python]
$\ProveT(P, \Pinstr, L, \traceTerm)$:
  rfset = $\{\}$
  $\tracesLoop$ = $\{\}$
  while (True):
    new_rfset = $\InferRF(\traceTerm, L)$
    rfset = rfset $\cup$ new_rfset|\label{progln:updaterfs}|
    if ($\IsUnchanged$(rfset):
      return ($\Unk, \tracesLoop$)
    else:
      cex = $\ValidateRFs$($P$, rfset)
      if (not cex):
        return ($
        , \{\}$)
      else:
        inps = $\GuessInputs$($\Pinstr$, cex)
        $\trace$ = $\Execute$($\Pinstr$, inps)
        $\trace_L$ = $\Project$($\trace, L$)
        $\traceBase, \traceTerm, \traceLoop$ = $\Partition(\trace_L, L)$
        $\tracesLoop$ = $\tracesLoop \cup \traceLoop$
\end{lstlisting}
\end{minipage}
\hspace{0.04\textwidth}
\begin{minipage}{0.47\textwidth}
\begin{lstlisting}[language=Python]
$\InferRF(\traceTerm, L)$:
  tcTrans = $\{\}$
  for $\tau_{L}$ in $\traceTerm$:
    tcTrans = tcTrans $\cup$ $\GenTCTrans$($\tau_{L}$)
  rfTemplate = $\GenRFTemplate(L)$
  rfset = $\{\}$
  while not $\IsEmpty$(tcTrans):
    ($s_1$, $s_2$) = $\RandPop$(tcTrans)|\label{progln:pop}|
    $t_1$ = rfTemplate($s_1$)
    $t_2$ = rfTemplate($s_2$)
    rf = $\Solve$(rfTemplate, $\{t_1>t_2, t_1\geq 0\}$)
    rfset = rfset $\cup$ $\{$rf$\}$
    tcTrans.filter(t: $\NotSatisfied$(t, rf))|\label{progln:filter}|
  return rfset
\end{lstlisting}
\end{minipage}
\end{tabular}
\caption{\label{fig:t-alg} Algorithm {\ProveT} for proving Termination, aided by dynamic inference of candidate ranking functions.}
\end{figure*}

% \begin{figure*}\centering
% 	\small
% \begin{tabular}{ll}
% \begin{minipage}{0.5\textwidth}
% \begin{program}[style=sf]
% Pr\tab oveT(P):
% 	$P_{instr}$ = instrument($P$)
% 	inps = GenRandomInps($P$)
% 	% ranking\_function\_set = []
% 	rfset = \{\ \}
% 	wh\tab ile (True):
% 		$\pi$ = Execute(P, inps)
% 		$\pi_{base}$, $\pi_{term}$, $\pi_{mayloop}$ = Partition($\pi$)
% 		new\_rfset = InferRF($P_{instr}$, $\pi_{term}$)
% 		rfset = rfset $\cup$ new\_rfset
% 		if\tab\ (rfset.unchanged):
% 			break
% 			\untab
% 		el\tab se:
% 			cex = validateRFs($P$, rfset)
% 			if\tab\ (not cex):
% 				return (True, rfset)
% 			\untab
% 			el\tab se:
% 				inps = GuessInput(cex)
% 			\untab
% 			\untab
% 	return (None, rfset)
% \end{program}
% \end{minipage}
% &
% \begin{minipage}{0.5\textwidth}
% \begin{program}[style=sf]
% In\tab ferRF($P_{instr}$, $\pi_{term}$):
% 	% rfTemplate = GenRFTemplate($P_{instr}$)
% 	tcTrans = \{\ \}
% 	fo\tab reach snap\_shot in $\pi_{term}$:
% 		tcTrans = tcTrans $\cup$ GenTCTrans(snap\_shot)
% 		\untab
% 	rfset = \{\}
% 	wh\tab ile tcTrans is non-empty:
% 		($s_1$, $s_2$) = RandPop(tcTrans)
% 		$t_1$ = rfTemplate($s_1$)
% 		$t_2$ = rfTemplate($s_2$)
% 		rf = Solve(rfTemplate, \{$t_1>t_2$, $t_1\geq 0$\})
% 		rfset = rfset $\cup$ \{rf\}
% 		tcTrans.filter(t: NotSatisfied(t, rf))
% 		\untab
% 	return rfset
% \end{program}
% \end{minipage}
% \end{tabular}
% \caption{\label{fig:t-alg} Algorithm \textsf{ProveNT} for proving Non-termination, aided by dynamic inference of candidate recurrent sets.}
% \end{figure*}

The algorithm for proving termination is summarized as follows, and two of the main subprocedures involved, \ProveT\ and \InferRF, are shown in Fig.~\ref{fig:t-alg}.

\paragraph{\ProveT} The procedure \ProveT\ aims to prove the termination of a loop $L$ in an instrumented program $\Pinstr$\ignore{ whose original version is $P$,} by inferring a set of ranking functions from a given set of terminating traces $\traceTerm$. (We discuss how $\Pinstr$ is built from $P$ in Section~\ref{sec:overview} and formalize it in Section~\ref{sec:impl}.) The procedure returns either the result \Term\ when the termination proof is successful  or otherwise, returns \Unk\ with a set of ``possibly non-terminating'' traces $\tracesLoop$ as a counterexample. Initially, the counterexample $\tracesLoop$ and the set of ranking functions \textsf{rfset} are initialised to be empty. The procedure then enters a loop until a valid set of ranking functions is found or until no progress is made when updating the set of ranking functions. Starting with the set of terminating traces $\traceTerm$, the subprocedure \InferRF\ is called to produce a set of ranking functions that attempts to cover those traces in $\traceTerm$. The details for this subprocedure is given in the next paragraph. The current set of ranking functions \textsf{rfset} is updated to include the resulting set of ranking functions (\textsf{new\_rfset}) from \InferRF. The loop in \ProveT\ exits if no new ranking functions were added. Otherwise, the updated set of ranking functions \textsf{rfset} is validated against the original program $P$ via a reachability prover (as is standard~\cite{Cook2006}).\ignore{such as Ultimate or CPAchecker chanh To explain how the ranking function set is validated.} If the prover returns no counterexample, which means the validation is successful, \ProveT\ returns \Term\ indicating that the loop $L$ is terminating (via the set of ranking functions \textsf{rfset}). On the other hand, if a counterexample to the set of ranking functions is found, then a new set of inputs is generated. The given program is executed on those new inputs and a set of concrete traces from these executions ($\pi$) is collected. These traces are then projected into the locations of interest in the loop $L$. That is, for each trace $\tau \in \pi$, the projection returns a sequence of states $\tau_{L}$, comprising the state right before the loop, the states 
reached inside the loop, right after the loop header, and the state at the loop's exit. Subsequently, the set of these sub-traces ($\pi_{L}$) are partitioned on whether they never enter the loop's body ($\traceBase$), whether they terminate ($\traceTerm$), and whether they reach the instrumented bound of iterations before terminating, and as such are classified as ``possibly non-terminating'' ($\traceLoop$). Finally, the traces in $\traceLoop$ are added into the counterexample $\tracesLoop$ and the procedure repeats the above steps with the new set of terminating traces $\traceTerm$.

\paragraph{\InferRF} This sub-procedure first generates a random sample of pairs of snapshots from the transitive closure of the concrete transition relation as follows.  For each terminating trace $\tau_{L} \in \traceTerm$, with an implicit order of appearance in the trace $\tau_{L}$ present, all combinations $(\sigma_1,\sigma_2)$ of the states $\sigma_1,\sigma_2\in \tau_{L}$ are generated, restricted so that $\sigma_1$ appears before $\sigma_2$ in $\tau_{L}$. The set of these combinations is randomly shuffled into a list, and the first $K$ pairs are selected, with $K$ being a predefined value for the desired size of the sample. All these samples from each trace $\tau_{L}$ are aggregated into the set \textsf{tcTrans}. The subprocedure \InferRF\ also generated a ranking function template, which is of the form $u_0 + u_1 \cdot v_1 + u_2 \cdot v_2 + \ldots u_n \cdot v_n$ for the set of variables $\{v_1,\ldots,v_n\}$ in the loop $L$ and the unknown coefficients  $u_0,u_1\ldots,u_n$. 

While the set \textsf{tcTrans} is non-empty, an element $(s_1,s_2)$ is randomly popped, and two instances $t_1,t_2$ of the template are produced for the two respective states $s_1$ and $s_2$. Given the valuation $\{h^i_1,\ldots,h^i_n\}$ of the set of variables $\{v_1,\ldots,v_n\}$ in the state $s_i$, for $i\in\{1,2\}$, the instance $t_i$ is of the form $u_0 + u_1\cdot h^i_1+u_2\cdot h^i_2+\ldots u_n\cdot h^i_n$. The solver from Z3 is then asked to return values for $u_0,\ldots,u_n$ that satisfy the constraints
\[\begin{array}{lcl}
		u_0+\sum_{1\leq j\leq n}u_j \cdot h^1_j > u_0+\sum_{1\leq j\leq n}u_j \cdot h^2_j,
		& \;\;\;\textrm{ and }\;\;\;&
		u_0+\sum_{1\leq j\leq n}u_j \cdot h^1_j \geq 0,
	\end{array}\]
while minimizing the value of $\sum_{0\leq j\leq n}|u_j|$. The resulting solution of values for $u_0,\ldots,u_n$ is added as a candidate ranking function to the set \textsf{rfset} of accumulated ranking functions. Any pair of states $(s_1,s_2)$ from the random sample of trasitive closure of the transition relation that was constructed earlier, that satisfies the latter candidate ranking function is removed from that sample, and the procedure continues with the remaining ones.

\paragraph{Correctness.} Sub-procedure \InferRF\ terminates since at each iteration of the loop we remove at least one of the pairs $(s_1,s_2)$ from $\textsf{tcTrans}$ (line~\ref{progln:pop} of \InferRF), but possibly more (line~\ref{progln:filter} of \InferRF). By construction, each ranking function returned by \InferRF\ handles at least one of the pairs $(s_1,s_2)$ in $\textsf{tcTrans}$. Such a candidate ranking function is only returned by \ProveT\ if it is validated on $P$ by a reachability solver.
On the other hand, it is not guaranteed that \ProveT\ will terminate. Because the traces are dynamically generated, and because the transitive closure is sampled randomly, a newly inferred candidate ranking function could potentially only handle few of the possible pairs of states in the actual transitive closure of the loop body. As a result, a new ranking function may be added to the set of possible ranking functions continuously (see line~\ref{progln:updaterfs} of \ProveT). 
%!TEX root = ./paper.tex

\section{Inferring Recurrent Sets for Non-Termination}\label{sec:alg_nt}

\begin{figure*}\centering
\begin{tabular}{ll}
\hspace{0.02\textwidth}
\begin{minipage}{0.47\textwidth}
\begin{lstlisting}[language=Python]
$\ProveNT(\Pinstr, L, \traceLoop)$:
  ($\Tstem$, $\Cloop$, $\Tloop$) = GetLoopInfo($\Pinstr$, $L$)
  $\Cmayloop$ = $\DInfer$($\traceLoop$)
  # stack of candidate recurrent sets
  stack S = $\{ (0, \Cloop), (0, \Cmayloop) \}$
  $\tracesTerm$ = $\{\}$
  
  while not IsEmpty(S):
    (depth, $R$) = Pop(S)
    if (depth>UPPERBOUND or $R(\Vars) {\centernot\implies} \Cloop(\Vars)$):|\label{progln:incloop}|
      continue;
    if $\IsValid(R(\Vars) \wedge \Tloop(\Vars, \Vars') \implies R(\Vars'))$:|\label{progln:validrs}|
      if $\IsSat(\Tstem(\Vars_0, \Vars) \wedge R(\Vars))$:|\label{progln:reachrs}|
        return ($\NonTerm, \{\}$)
    else:
      $RS, \traceTerm$ = $\text{\RefineRS}(R, \Pinstr, L, \Tstem, \Tloop)$
      $\tracesTerm$ = $\tracesTerm \cup \traceTerm$
      for $R'$ in $RS$: 
        S.push($(\text{depth}+1, R')$)
  return ($\Unk, \tracesTerm$)
\end{lstlisting}
\end{minipage}
\hspace{0.04\textwidth}
\begin{minipage}{0.47\textwidth}
\begin{lstlisting}[language=Python]
$\RefineRS(R, \Pinstr, L, \Tstem, \Tloop)$:
  $R \text{ as } \bigwedge_{i}R_i$
  $RS = \{\}$
  $\tracesTerm$ = $\{\}$
  for $R_i$ in $R$:
    $r_i = (R(\Vars) \wedge \Tloop(\Vars, \Vars') \implies R_i(\Vars'))$
    if $\IsSat(\neg r_i)$:
      inps = $\GuessInputs(\Tstem(\Vars_0, \Vars) \wedge \neg r_i(\Vars, \Vars')$
      $\trace$ = $\Execute$($\Pinstr$, inps)
      $\trace_L$ = $\Project(\trace, L)$
      $\traceBase, \traceTerm, \traceLoop$ = $\Partition$($\trace_L, L$)
      $\Cterm$ = $\DInfer(\traceTerm)$
      $\Cmayloop$ = $\DInfer(\traceLoop)$
	    $\tracesTerm$ = $\tracesTerm \cup \traceTerm$
      $\Cterm \text{ as } \bigwedge_i \Cond_{i}$
      for $\Cond_i$ in $\Cterm$:
        $RS = RS \cup \{R \wedge \neg \Cond_{i} \}$
      $RS = RS \cup \{ \Cmayloop \}$
	return $(RS, \tracesTerm)$
\end{lstlisting}
\end{minipage}
\end{tabular}
\caption{\label{fig:nt-alg} Algorithm \ProveNT\ for proving Non-termination, aided by dynamic inference of recurrent sets.}
\end{figure*}

The algorithm \ProveNT\ for proving non-termination is given in Fig.~\ref{fig:nt-alg}. 
The input is an instrumented program $\Pinstr$, the loop $L$ currently being analysed, and a set $\traceLoop$ of traces that may be non-terminating. The procedure outputs either that a recurrent set was found (\NonTerm), or that such a recurrent set was not found (\Unk) together with a set of traces that were found to be terminating. 
% output is either success a collection
% of \emph{preconditions} for recurrent sets $\rcrSet$.
% If $\rcrSet$ is empty, then we have failed to prove
% non-termination. If the disjunction of all conditions in $\rcrSet$ is true, then $P$ always non-terminates\timos{better `never terminates'?}.
%
The algorithm is aided by a dynamic sub-procedure \RefineRS\ for
guessing candidate recurrent sets, which are then validated.

\begin{wrapfigure}[9]{r}[34pt]{0.6\columnwidth}
\includegraphics[width=0.51\columnwidth]{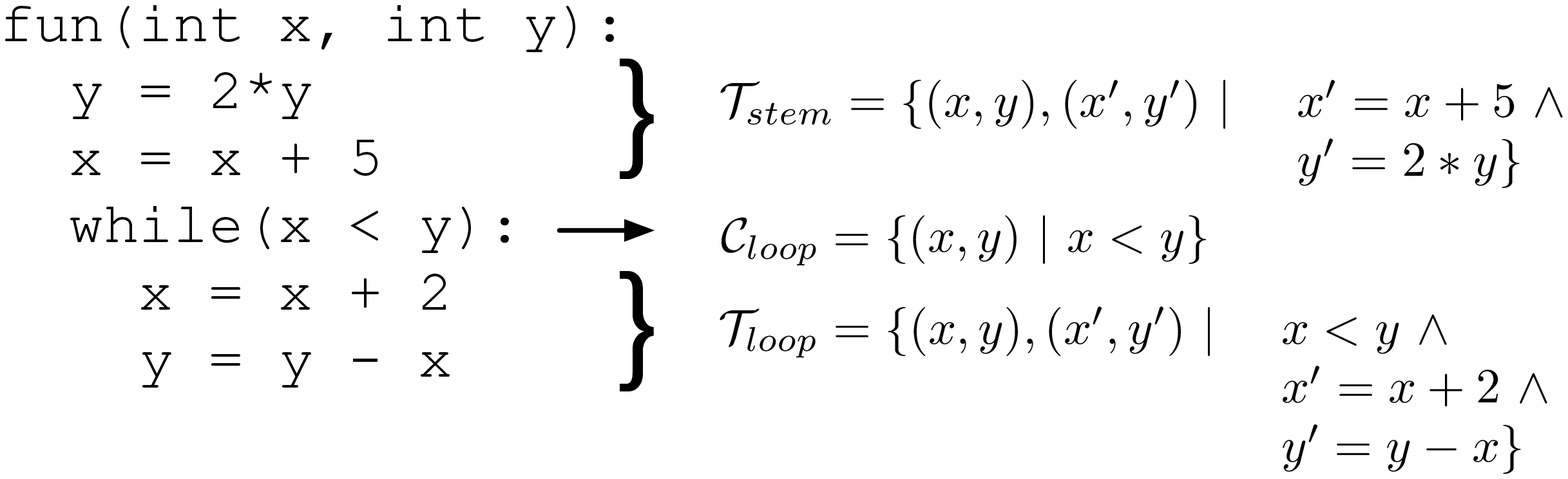}
\vspace{-0.2cm}
\caption{Illustration of $\Tstem$, $\Cloop$ and $\Tloop$}\label{fig:loopsummary}
\end{wrapfigure}

\ProveNT\ begins by collecting summaries for the loop $L$ in $P$. We use standard techniques~\cite{Cook2006} to represent $L$ in terms of three entities:
\begin{itemize}
\item $\Tstem$ is a state relation that over-approximates the transition from the entry point of the program up to the entry point of loop $L$.
\item $\Cloop$ is a state predicate that over-approximates the condition for entering the loop.
\item $\Tloop$ is a state relation that over-approximates all transitions from the beginning of the body of the loop, back to the loop header.
\end{itemize}
% \red{maybe diagram of these.}\timos{see figure}
An illustration of these entities is given in Fig.~\ref{fig:loopsummary}.
%\begin{figure*}
%\includegraphics[width=3.5in]{stem-loop.pdf}
%\caption{Illustration of $\Tstem$, $\Cloop$ and $\Tloop$}\label{fig:loopsummary}
%\end{figure*}
The algorithm is structured using a stack \textsf{S} as a work list, tracking candidate recurrent sets that will later be examined and possibly refined. To begin with,  
we ambitiously select $\Cloop$ to be the first candidate recurrent set. The stack element also includes an integer 0, to track the exploration depth, so that we can later bound the search. We also add the condition $\Cmayloop$ into the work list \textsf{S}, which is dynamically inferred from the set $\traceLoop$ of possibly non-terminating traces received from a failed termination proof. 
% Finally, we initialize the output $\rcrSet$ to be empty and instrument the program~\cite{subsec:pinstr} so that we can collect executions.

The main loop iterates as long as \textsf{S} is non-empty and no valid recurrent set was found. Popping a candidate $R$ off the stack, if we have gone beyond some upper bound, then we simply ignore $R$ rather than exploring further refinements of $R$. $R$ is also ignored if it doesn't even imply the loop condition: it could not be a recurrent set.
We next use an SMT query \IsValid\ to check whether $R$ is indeed a recurrent set, \ie, Definition~\ref{def:rcr}: if $R$ holds of variables $\Vars$, and a loop body transition to $\Vars'$ is possible, then $R$ must hold of $\Vars'$.
If $R$ is a recurrent set, then we check that at least some state in $R$ is reachable from an initial state, using $\Tstem$, and if it is we have succeeded in proving the program to be non-terminating.
% we add the \emph{precondition} of $R$ to $\rcrSet$: the condition on the initial states that would lead execution to this $i$th loop and a state that is in recurrent set $R$.
%
Alternatively, if $R$ is not a recurrent set, we explore further by refining $R$, with respect to this loop, using subprocedure \RefineRS\ discussed below. This subprocedure also collects any terminating traces found during the refinement. Such terminating traces are evidence that the program is terminating, and thus useful for the case where non-termination fails to be proved and the algorithm switches to proving termination.

%!TEX root = ./paper.tex

\paragraph{\RefineRS}
This subprocedure, shown in Fig.~\ref{fig:nt-alg}, takes as input the current candidate recurrent set $R$ with respect to the loop $L$ and returns a set of new candidate recurrent sets $RS$. The input recurrent set $R$ is assumed to be a conjunction of $\{R_i\}_{i\leq k}$, for some $k\in\mathbb{N}$, and it is known that $R$ is not a recurrent set for the transition relation $\Tloop$. As such there are two states $\sigma$ and $\sigma'$ (\ie\ two valuations of the set of variables $\Vars$), for which the formula on the left does not hold:
\[\begin{array}{ccc}
\bigwedge_{i\leq k}R_i(\sigma)\land\Tloop(\sigma,\sigma') \implies \bigwedge_{i\leq k}R_i(\sigma')
& \hfill &
	r_i = \bigwedge_{i\leq k}R_i(\sigma)\land\Tloop(\sigma,\sigma')\implies R_i(\sigma')
\end{array}\]
%\begin{equation}
%\end{equation}
Therefore, for at least one of the $i\leq k$, the formula on the right does not hold.
The candidate recurrent set $R$ is updated for each such $R_i$ as described next. The algorithm proceeds via $\GuessInputs{}$ using SMT to generate solutions $\Vars_0$ to the formula $\exists\Vars,\Vars'\ \Tstem(\Vars_0, \Vars) \wedge \neg r_i(\Vars, \Vars')$, which are used as inputs to execute $\Pinstr$. Informally, these inputs are witnesses to a path, via $\Tstem$, from the initial state to a state on which the recurrent set fails.
After the program is executed using the resulting inputs $\textts{inps}$ and a set of traces is produced as a result. The traces are first projected---to include the instrumented information regarding only the loop $L$ being analyzed---and then partitioned into: the traces $\traceBase$ that never enter the loop, traces $\traceTerm$ that definitely terminate, and traces $\traceLoop$ that may be non-terminating, as the execution for the latter reached the imposed loop bound.
It should be noted that, since any candidate $R$ implies $\Cloop$, if the program is deterministic, and assuming soundness of the preceding subprocedures, then the inputs $\textts{inps}$ will not cause any traces that never enter the loop to be generated, and thus $\traceBase$ will be empty.
Traces $\traceTerm$ are used to dynamically infer a condition $\Cterm$ that captures the set of states reached right after the loop header by those terminating traces and a similar condition $\Cmayloop$ is inferred using $\traceLoop$. The accumulating set $\tracesTerm$ is updated to include $\traceTerm$ and the recurrent set $RS$ is then updated as follows.
For every conjunct $\Cond_i$ of $\Cterm$, $RS$ is updated to include the strengthened candidate recurrent set $R\land\lnot\Cond_i$ in which any states in $\Cond_i$ that is possibly in a terminating trace is excluded from the candidate. The condition $\Cmayloop$ is also included as a new candidate recurrent set since it captures all states that are in possibly non-terminating traces. At the end, the procedure $\RefineRS$ returns the set of candidate recurrent sets constructed, together with any terminating traces accumulated in $\tracesTerm$.
% If the cardinality of the terminating traces (\ie\ the set of traces $\traces_{base}\cup\traces_{term}$) is much larger than the possibly non-terminating set of traces (\ie\ $\traces_{mayloop}$), then the given conjunct $R_i$ is replaced with the negation of the condition $\Cond_{term}$ and the resulting formula is added to the new set of possible recurrent sets to be returned by $\Refine$. Otherwise, the condition $\Cond_{mayloop}$ is added to the new set of possible recurrent sets.

Consider the example to the right with non-linear expressions to illustrate how \ProveNT\ and
 \RefineRS\ works. The summary of this loop is: $\Tstem = \true$, $\Cloop = t \leq n^2 + 1$, and
\begin{wrapfigure}[6]{r}[34pt]{5.1cm}
\begin{lstlisting}[language=Python,style=pnnstyle]
    int t, n, m
    while (t <= n*n + 1):
      t = t + 2*m
      n = n + 1
\end{lstlisting}
\end{wrapfigure} 
$\Tloop = t \leq n^2 + 1 \wedge t' = t + 2m \wedge n' = n + 1 \wedge m' = m$.
The procedure \ProveNT\ first uses the loop condition $t \leq n^2 + 1$ as a candidate recurrent set and checks if the implication
$t \leq n^2 + 1 \wedge \Tloop \implies t' \leq n'^2 + 1$ is valid. As this is not the case, \ProveNT\ invokes  \RefineRS\ on this invalid recurrent set to refine it. Then \RefineRS\ finds a set of inputs over the variables $(t, n, m)$ that invalidate the implication, such as $\{(29, -6, -1), (13, -4, 0), (1, 0, 1), (1, -1, 2), (0, 0, 3), (1, 1, 4), \ldots\}$. The program execution over these inputs produces only terminating traces and a dynamic invariant inference tool, like Dig~\cite{tosem2013,nguyen2012using}, can generate the condition $m \geq -1$ from the snapshots at the beginning of the loop's body in those traces. This possibly terminating condition (see $\Cterm$ in \RefineRS) is used to refine the current candidate into a new one $t \leq n^2 + 1 \wedge \neg(m \geq -1)$. Unfortunately, this new candidate recurrent set is still invalid and \RefineRS\ generates a new set of inputs from its validity check, that is $\{\ldots, (78, -9, -5), (26, -5, -4), (24, -5, -3), (15, -4, -2)\}$. Again, all these inputs lead to terminating traces, from which a new condition $n \leq m - 1 \wedge m \leq -2$ can be dynamically inferred. From this new condition, \RefineRS\ returns two new candidate recurrent sets by strengthening the current candidate $t \leq n^2 + 1 \wedge \neg(m \geq -1)$:
\begin{enumerate}
	\item $t \leq n^2 + 1 \wedge \neg(m \geq -1) \wedge \neg(n \leq m - 1)$, and
	\item $t \leq n^2 + 1 \wedge \neg(m \geq -1) \wedge \neg(m \leq -2)$.
\end{enumerate}
% $t \leq n^2 + 1 \wedge \neg(m \geq -1) \wedge \neg(n \leq m - 1)$ and $t \leq n^2 + 1 \wedge \neg(m \geq -1) \wedge \neg(m \leq -2)$.
Finally, the procedure \ProveNT\ determines that the new candidate (1) is a valid recurrent set and returns the result \NonTerm.

% \commentout{
% $ \GetModels(\phi)$ : (standard)\\
% $ \GuessInputs{}_S(\cex, \Tstem) = \{ \vln \mid \red{fix.}\Tstem(\vln,\cex)\}$\\
% $ \GuessInputs{}_D(\cex, path_{stem}) = \{ \vln \mid \exists \trace. \trace(\vln,\cex,path_{stem})\}$\\
% $ \GetTraces{}(P, \vln_0) : \{ \trace \mid \vln_0 \Rightarrow \trace^0 \}$\\
%
% $ \Partition{} : \traces \rightarrow \traces \times \traces \times \traces$ such that...\\
% }

%The recurrent sets $\rcrSet$ returned by \textsf{InferRS} may not be sound: they are synthesized based on inference from samples of dynamic traces. Therefore, the next step is to filter $\rcrSet$ \emph{prove} 

%\paragraph{\Precond}
%We remind the user that the relation $\Tstem$ relates the initial states of the given program $P$ to the states reached right before entering the loop at hand. Given $\Tstem$ and a recurrent set $R$, then $\Precond$ is define to be the formula $\exists \bar{X}'.\ \Tstem(\bar{X},\bar{X}')\land R(\bar{X}')$, with free variables $X$. In other words, the precondition for a recurrent set $R$ is the set of states $S$ for which there exists another set of states $S'$, reacheable from $S$ via the relation $\Tstem$, and where $S'$ satsifies the recurrent set.\timos{not needed right?}

\paragraph{\DInfer.}
We use dynamic invariant inference to guess conditions from terminating traces and potentially non-terminating traces which are then used to refine the invalid candidate recurrent sets.
Dynamic invariant generation works, pioneered by the tool Daikon~\cite{ernst2007daikon,ernst2001dynamically}, learns candidate invariants from program execution traces and templates (e.g., equalities, inequalities).
Recent works in dynamic invariant generation are capable of generating very expressive invariants (e.g., nonlinear invariants).
In addition, many works integrate dynamic inference with static checking to remove spurious invariants.
We use the DIG~\cite{nguyen2012using,tosem2013} tool to infer numerical invariants from traces.
DIG supports nonlinear equations as well as several other forms of inequalities such as octagonal invariants and max/min-plus invariants.
DIG reduces the problem of nonlinear equation solving to linear equation and ussing terms to represent nonlinear polynomials and uses linear constraint solving to find octagonal and max/min-plus invariants.
In addition, DIG implements a counterexample-guided invariant generation technique that iteratively infer candidate invariants from program traces and check them using symbolic execution, which, for incorrect invariants, returns counterexamples that are used as traces to help DIG infer better results in the next iteration.

%\begin{itemize}
%	\item $\GetModels(\phi)\subseteq \Sigma\times\Sigma$ (\ie\ sets of valuations for $V$ and $V'$)
%	\item $\GuessInputs{}_{\{S,D\}}(\cex, \Tstem)\subseteq \Sigma$
%	\item $\GetTraces(P_{instr}, \vln_0)\subseteq \Sigma^*$ (\ie\ sets of traces, or sets of sequences of states)
%	\item $\DInfer(\traces):$ a formula with free variables over $V$
%	\item $\Precond(R,P_{instr})$: a formula with free variables over $V$
%\end{itemize}

%!TEX root = ./paper.tex

\subsection{Correctness}
\paragraph{\ProveNT} For correctness of the algorithm \ProveNT\ we aim to show that if its output is $(\textts{NonTerm}, \{\})$ then there is at least one execution of the program $P$, that leads to a non-terminating execution of the loop $L$ at hand. For what follows, we assume that
\begin{enumerate}
	\item for any state that satisfies $\Cloop$, there is a valid transition from that state in the program $P$,
	\item $\Tstem$ is an exact representation, or at worst an under-approximation, of the transition from the entrypoint to the loop header, and
	\item $\Tloop$ is an exact representation, or at worst an over-approximation, of the loop body transition.
\end{enumerate}
 % and similarly , and which in the case that it is an under-approximation, it is also total on the loop condition $\Cloop$.\timos{Discuss here that our problem is over-approxiamtion that one one hand satisfies this totality condition but may lead to unsound results, thus we need witness generation.}

The procedure \ProveT\ will declare that the input loop is non-terminating only when both $\textts{IsValid}(R(\Vars) \wedge \Tloop(\Vars, \Vars') \implies R(\Vars'))$ and $\textts{IsSat}(\Tstem(\Vars_0, \Vars) \wedge R(\Vars))$ hold (see lines~\ref{progln:validrs} and~\ref{progln:reachrs} of \ProveNT). In other words,
$$(i)\hspace{5pt}\exists\Vars_0,\Vars\ \Tstem(\Vars_0, \Vars) \wedge R(\Vars)\hspace{20pt}\textrm{ and }\hspace{20pt}(ii)\hspace{5pt}\forall\Vars,\Vars' R(\Vars) \wedge \Tloop(\Vars, \Vars') \implies R(\Vars').$$
Formula $(ii)$, together with the assumption (3) above implies the condition (3) of Definition~\ref{def:rcr}.
% Suppose $\phi$ is one condition returned by \ProveNT. Then $\phi$ is of the form $\exists \bar{X}'.\ \Tstem(\bar{U},\bar{X}')\land R(\bar{X}')$, where $R$ is such that for all states $Q$ and $Q'$, $R(Q)\land\Tloop(Q,Q')\implies R(Q')$ (see line~\ref{progln:newrs} of \ProveNT) and as such, together with the assumption (3) above, the condition (3) of Definition~\ref{def:rcr} holds.
From formula (i) and the assumption (2) above, there is a state $S'$ at the loop header that can be reached from $S$, such that $R(S')$ holds which implies that condition (1) of Definition~\ref{def:rcr} holds for a reachable state in $P$ from $S$. Finally, given that $R$ implies $\Cloop$ (see line~\ref{progln:incloop} of \ProveNT), and given the assumption (1) above, it follows that the real transition relation for the loop is total on $R$. Therefore there is a non-terminating execution of $P$ starting from the state $S$.
%In our tool currently, $\Tstem$
We should note that, given that $\Tstem$ is an over-approximation in reality, our implementation could simply check if a witness path exists.

Further, the algorithm terminates, since whenever a new candidate recurrent set $R$ is added the variable $\textsf{depth}$ is increased and the recurrent sets with an accompanying $\textsf{depth}$ of value higher than $\textsf{UPPERBOUND}$ are ignored (see line~\ref{progln:incloop} of \ProveNT).

\section{An integrated algorithm}
\label{sec:tnt}

We now describe \ProveTNT, an algorithm supported with dynamic analysis, that mixes termination and non-termination reasoning, allowing the failed outcomes of one endeavor to provide feedback to 
 the other. In this algorithm, \ProveNT\ consumes the previously ignored argument $\pil$ 
%
%\hspace{0.1\textwidth}
%\begin{minipage}{0.35\textwidth}
%	\begin{lstlisting}[language=Python,style=pnnstyle]
%	while(x>=0):
%	  x = x + y
%	\end{lstlisting}
%	\vspace{-20pt}
%	\caption{\label{fig:ex-tnt}}
%	\vspace{60pt}
%	\begin{lstlisting}[language=Python,style=pnnstyle]
%	while(x<1000):
%	  x = x + 1
%	\end{lstlisting}
%	\vspace{-20pt}\red{move to inline}
%	\caption{\label{fig:ex-tnt-2}}
%\end{minipage}
% \caption{\label{fig:integratedalg} The integrated algorithm for approving termination and non-termination, via mutual feedback.}
%
% \begin{wrapfigure}[17]{r}[34pt]{0.55\columnwidth}
\begin{figure}
% \begin{minipage}{0.48\textwidth}
%\hspace{0.4in}
\begin{lstlisting}[language=Python,xleftmargin=2.0ex,basicstyle=\ttfamily\footnotesize]]
$\ProveTNT$($P$):
  $\Pinstr$ = Instrument($P$)
  inps = GenRandomInputs($\Pinstr$)
  $\pi$ = $\Execute$($\Pinstr$, inps)
  $\mathcal{L}$ = $\GetLoopSeq$($\Pinstr$)
  
  for $L$ in $\mathcal{L}$:
    $\trace_L$ = $\Project$($\trace, L$)|\label{progln:project}|
    $\traceBase, \traceTerm, \traceLoop$ = Partition($\trace_L, L$)
    if $\card{\traceLoop}$ >> $\card{\traceBase \cup \traceTerm}$:
      $r_{nt}, \tracesTerm$ = $\ProveNT$($\Pinstr, L, \traceLoop$)
      if $r_{nt}$ is $\NonTerm$:
        return $\NonTerm$
      else:
        $r_{t}, \_$ = $\ProveT$($P, \Pinstr, L, \traceTerm \cup \tracesTerm$)
        if $r_{t}$ is $\Unk$:
          return $\Unk$
    else:
      $r_{t}, \tracesLoop$ = $\ProveT$($P, \Pinstr, L, \traceTerm$)
      if $r_{t}$ is $\Unk$:
        $r_{nt}, \_$ = $\ProveNT$($\Pinstr, L, \traceLoop {\cup} \tracesLoop$)
        if $r_{nt}$ is $\NonTerm$:
          return $\NonTerm$
        else:
          return $\Unk$
  return Term
\end{lstlisting}
\caption{\label{fig:integratedalg} The integrated algorithm for approving termination and non-termination, via mutual feedback.}
% \end{minipage}
% \end{wrapfigure}
\end{figure}
(a set of potentially non-terminating traces returned by \ProveT) and \ProveT\ consumes the previously ignored argument $\pit$ (a set of terminating traces from \ProveNT).

The procedure \ProveTNT\ is given in Fig.~\ref{fig:integratedalg} and begins by instrumenting the input program $P$, generating random initial inputs, and executing the instrumented program on those inputs to get a set $\pi$ of concrete traces.
These executions may be used for reasoning about multiple loops in the program, and avoid the need for re-execution. We
then iterate over the loops in the program in a post-order 
fashion, in  which the top-down innermost loop will be analyzed first. If that loop is proved to be non-terminating, the procedure returns the result \NonTerm\ immediately. Otherwise, it continues to analyze the next loop in the post-order sequence. At the end, the procedure returns the result \Term\ when all loops in the program are proved to be terminating.

Within each loop $L$ we project on the set of traces, focusing on only those that reach $L$'s header and keeping only the relevant snapshots from the instrumentation on that loop in $\pi_{L}$ (see line~\ref{progln:project} in \ProveTNT). We then partition $\pi_{L}$ into the three classes of traces $\pib,\pit,\pil$, similarly to what was described in previous sections. We next make a decision as to whether we should attempt non-termination or termination reasoning first.
Our algorithm heuristically chooses which action to perform after comparing the sizes of terminating trace sets ($\pib$ and $\pit$) and the potentially non-terminating one ($\pil$). In our implementation, we decide to prove non-termination first when the number of the potentially non-terminating traces is four times larger than the total size of terminating traces.
In the case the algorithm succeeds in proving the chosen analysis, it moves to the next step as described above (\ie\ returning \NonTerm\ immediately if \ProveNT\ is chosen or analyzing the next loop if \ProveT\ is chosen). Otherwise, the chosen sub-procedure will return counterexamples in the form of new traces, that can be used, together with the traces collected from running the random inputs, as input to the alternative analysis.

Consider the simple program: \lstinline|while(x>=0): x = x + y|.
%example in Fig.~\ref{fig:ex-tnt}.
This example conditionally terminates, depending on the initial values of \textts{x} and \textts{y}. 
That is, the loop does not terminate when $\textts{x} {\geq} 0$ and $\textts{y} {\geq} 0$ and terminates otherwise.
% \begin{wrapfigure}[4]{r}[34pt]{0.3\columnwidth}
% \begin{quote}
% % \lstset{}
% \begin{lstlisting}[language=Python,style=pnnstyle]
% while(x>=0):
%   x = x + y
% \end{lstlisting}
% \end{quote}
% \caption{}\label{fig:ex-tnt}
% \end{wrapfigure}
Given that the random inputs are evenly-distributed then \textts{x} is negative in roughly half of the random inputs, on which the loop terminates. From the heuristic for choosing the sub-procedure, \ProveTNT\ may decide to attempt proving termination first. More specifically, in our implementation, we decide to prove non-termination first when the number of the potentially non-terminating traces is four times larger than the total size of terminating traces. The sub-procedure \ProveT\ may find a ranking function such as \textts{x} from the terminating traces. However, it is not a valid ranking function for all inputs and the validation in \ProveT\ returns counterexamples whose corresponding inputs create potentially non-terminating execution traces, such as $[(\textts{x}{=}0, \textts{y}{=}0), (\textts{x}{=}0, \textts{y}{=}0), \ldots]$, $[(\textts{x}{=}3, \textts{y}{=}1), (\textts{x}{=}4, \textts{y}{=}1), \ldots]$, etc., in which the ranking function \textts{x} is not decreasing. Since there is no terminating trace generated from those inputs, \ProveT\ gives up and returns such potentially non-terminating execution traces as its counterexample traces. At this point, our \ProveTNT\ algorithm switches gears and uses these counterexample traces as inputs to \ProveNT. Finally, \ProveNT\ proceeds on these traces and finds a recurrent set $\textts{x} {\geq} 0 \wedge \textts{y} {\geq} 0$ from them to confirm the loop's non-termination.

The proving strategy in \ProveTNT\ also helps to overcome the scenario when execution traces from a terminating program, like the simple loop in 
this program:
\lstinline|	while(x<1000): x = x + 1|. This is wrongly categorized as potentially non-terminating due to the predefined instrumented execution bound (\eg\ \textts{500}) being reached before the loop terminates.
% \begin{wrapfigure}[4]{r}[34pt]{0.35\columnwidth}
% \begin{quote}
% \begin{lstlisting}[language=Python,style=pnnstyle]
% while(x<1000):
%   x = x + 1
% \end{lstlisting}
% \end{quote}
% \caption{}\label{fig:ex-tnt-2}
% \end{wrapfigure}
In this example, the execution traces with inputs where $\textts{x} < 500$ are considered as potentially non-terminating. Note that on those inputs, the collected traces from that loop are identical to traces collected from the non-terminating loop \textts{while True: x = x + 1}. If those inputs dominate the set of random inputs then the procedure \ProveTNT\ may attempt proving non-termination first.
The sub-procedure \ProveNT\ then starts with the first candidate recurrent set $\textts{x} {<} 1000$ and performs a check on it with the implication $\textts{x}{<}1000 \wedge \textts{x}'=\textts{x}{+}1 \implies \textts{x}'{<}1000$. The implication does not hold and there is only one input of $\textts{x}{=}999$ and the corresponding terminating trace $[\textts{x}{=}999, \textts{x}{=}1000]$ are generated from it as counterexample. Due to the lack of data, the dynamic inference is not triggered and there is no new candidate recurrent set generated. The procedure \ProveTNT\ then passes that terminating counterexample trace to the sub-procedure \ProveT\ for proving termination. From that trace, \ProveT\ can easily find the ranking function $999 - \textts{x}$ to prove the loop's termination. Interestingly, the \ProveT\ alone cannot prove the termination of this loop due to the lack of terminating traces. This happens since we usually prefer to generate small random inputs and limit the number of them, which may help to reduce the program execution time, for efficiency. In this example, we can try inputs larger than the predefined instrumented bound (\ie\ $\textts{x} \geq 500$) but the same problem may occur on other examples if the generated inputs are not large enough. For example, for the same program but with the loop condition replaced with \textts{x < 10000}, the algorithm would require some inputs where $\textts{x}$ would be larger than or equal to $9500$.
%!TEX root = ./paper.tex

\section{The {\tool} Tool}
\label{sec:impl}

We have realized our learning-based algorithms in a new tool called \tool.
\tool\ employs the power of several major existing tools, yet our particular combination of them allow \tool\  to do things that none of these tools can achieve individually (see Fig.~\ref{fig:tools}).
%   I also think we should create a diagram to show how Dynamo uses these tools.
We found that we were able to use these tools with few modifications and, consequently, our framework allows us to benefit from improvements in those tools or substitute alternatives. 
For SMT, we use SMTlib and for reachability, we follow the SV-COMP~\cite{svcomp2020} format.
% I also think we should emphasize that we use these tools as they are and they can be replaced easily,  e.g., instead of Z3 or CVC4,  the alg can use whatever specialize SMT solver available, as long as they support SMTcomp format.  Similarly,  instead of Cpachecker / Ultimate,  we can replace them with whatever support SVCOMP format
We now discuss some of the key components of the implementation.

We perform two transformations on the input program. For validating our guesses in termination reasoning, we use a standard transformation~\cite{Cook2006} that lets us input candidate rank functions and apply a reachability prover. This is a common technique and we have briefly described it in Sec.~\ref{sec:background}.
The second transformation (described in Section~\ref{sec:overview}) involves (i) instrumenting the program to collect states and traces and (ii) truncating potentially infinite loops. For a formal description of this transformation, see Appendix~\ref{apx:cfa}.

\ignore{
\paragraph{Integrated tools.}
Once we have performed the above instrumentation, the core algorithms of 
\begin{wrapfigure}[15]{r}[34pt]{0.6\columnwidth}
\includegraphics[width=2.7in]{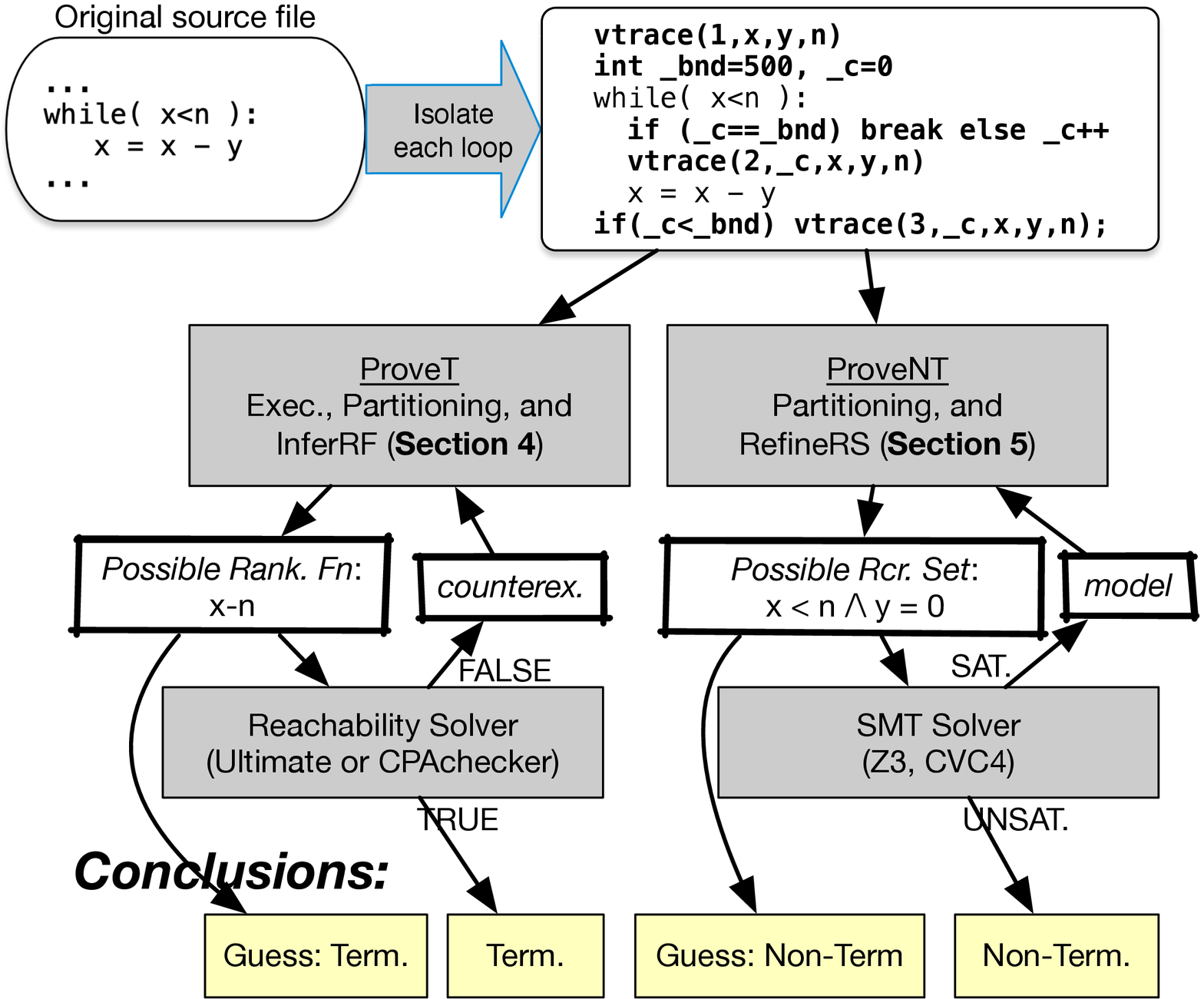}
\end{wrapfigure}
\tool\ begin.
The flows are depicted in the diagram to the right. \chanh{expand, explain diagram}
\paragraph{Sampling inputs and snapshots for efficiency.}
\begin{itemize}
  \item  Termination: we can do that due to disjunctive well-foundedness, ranking functions distribute in the transitive closure so we can still discover high-quality ranking functions from samples. Non-termination: sampling helps to quickly find conditions to strengthen the candidate recurrent sets and avoid overfitting on those conditions.
  \item Symbolic execution with a target location? Problems with other tools?
\end{itemize}
\paragraph{Dynamic Inferrence.}
}

%!TEX root = ./paper.tex

\newcommand\rTO{{\bf T.O.}}
\newcommand\UNK{{\bf ?}}
\newcommand\rUNK{{\bf ?}}
\newcommand\rTRUE{\checkmark}
\newcommand\rSCD{\checkmark}
\newcommand\rFALSE{$\chi$}
\newcommand\nparse{\red{prs}}
\newcommand\rAF{--} %\red{art}}
\newcommand\ULTIMATE[1]{#1}
\section{Evaluation}\label{sec:eval}

Our main goal of \tool\ is to improve the state-of-the-art in termination and non-termination reasoning to better support non-linear (NLA) programs. To this end, Sections~\ref{subsec:nlat} and~\ref{subsec:nlant} report experimental results on those programs. However, in Section~\ref{subsec:linear} we first evaluated \tool\ to see how it performs on linear programs, particularly in comparison with the state-of-the-art tool \UAutomizer\ from Ultimate~\citet{ultimate}, which is the winner of the Termination category in the recent Competition on Software Verification (SV-COMP) \cite{svcomp2020}.

%\red{if Ult won term-crafted-lit, we hsould say that to justify that we only compared against ult}

Our experiments were all run on a 20-core Intel(R) Core(TM) i7-6950X CPU @ 3.00GHz.
\tool\ in general take advantages of parallel processing when possible. For example, 
in termination reasoning, multiple instances of Ultimate's variants (Automizer and Taipan) and CPAchecker are invoked
to validate the termination results. In non-termination reasoning, we use these CPU cores to run the symbolic execution tool CIVL 
to obtain program information at multiple depths. The dynamic inference tool Dig also computes invariants simultaneously.
The timeout for each benchmark program is 400s.
 
\subsection{Linear programs}
\label{subsec:linear}

Although our main goal was to support NLA programs (\ie\ expressivity) we nonetheless compared our work against the state-of-the-art termination tool \UAutomizer. We ran both \UAutomizer\ and \tool\ on the 61 termination benchmarks and 5 non-termination benchmarks from the SV-COMP suite \textts{termination-crafted-lit}, which were used in SV-COMP 2020. Note that this folder contains other benchmarks that are for other properties like overflow.

\newcommand\TCLheader[1]{ & \multicolumn{1}{c|}{\tool} & \multicolumn{2}{c|}{{\bf Learning}}  & \multicolumn{2}{c|}{{\bf Validate}}  & \multicolumn{2}{c||}{\bf Total} & \multicolumn{2}{c|}{\UAutomizer}\\%
{\bf Benchmark} & {\bf Learned #1} & T(s) & Res &  T(s) & Res &  T(s) & $\sigma_5$ & T(s) & Res\\}

\newcommand\figtitle[1]{\begin{center}{\sc #1}\end{center}}

\begin{figure}\centering\footnotesize
\def\arraystretch{0.9}
\setlength{\tabcolsep}{0.3em}
\figtitle{\tool\ and Ultimate on LIN Termination Programs from SV-COMP}
\begin{tabular}{|p{1.6in}||p{1.5in}|rc|rc|rr||rc|}
\hline
\TCLheader{Rank Functions}
\hline
\input{termcraftlitsummary}
\multicolumn{10}{|c|}{}\\
\multicolumn{10}{|c|}{...}\\
\multicolumn{10}{|c|}{(Results of for the other 46 benchmarks in Appendix~\ref{apx:tcl}.}\\
\hline
\end{tabular}\\
%\textrm{AlDaFeGo-SAS2010-easy1}: &  -x, -x + 14\cdot z, -x + 18\cdot z, -x + 22\cdot z, -x + 30\cdot z, -x + -2\cdot z, -x + -3\cdot z, -x + -4\cdot z, \\
%&  -x + -5\cdot z, -x + -6\cdot z, -x + -17\cdot z, 25 + -x, 29 + -x, 36 + -x, 37 + -x, -x + -38\cdot z   \\
%\textrm{ChCoGuSaYa-ESOP2008-easy1}:  &  -x + -z, -x + 3\cdot z, -x + 4\cdot z, -x + -2\cdot z, -x + -3\cdot z, -x + -4\cdot z, -x + -5\cdot z, -x + -6\cdot z, -x + -9\cdot z, -x + -22\cdot z, -x + 5\cdot z, -x + 7\cdot z, -x + 15\cdot z, -x + 16\cdot z, 15 + -x, 21 + -x, 25 + -x, 34 + -x, -x + 36\cdot z \\
\footnotesize
\[\begin{array}{lll}
\textrm{KrShTsWi-CAV2010-Fig1.c:} & 28{-}x, 82{-}x, 88{-}x, 90{-}x, 104{-}x, 118{-}x, 144{-}x, 156{-}x,\\
 & 212{-}x, 214{-}x, 228{-}x, 234{-}x, 246{-}x  \\
\textrm{Urban-WST2013-Fig2.c:} & -x1, -x1 + 5{\cdot}x2, -x1 + 6{\cdot}x2, -x1 + 7{\cdot}x2, -x1 + 8{\cdot}x2, -x1 + 9{\cdot}x2, -x1 + 10{\cdot}x2, -x2 
\end{array}\]
\caption{\label{fig:termlin} Results of applying \UAutomizer\ and \tool\ on the 61 termination benchmarks from SV-COMP \textts{termination-crafted-lit}. For lack of space, we only show 15 rows (every 4th row); the rest are shown in Appendix~\ref{apx:tcl}. Names are abbreviated; see Appendix~\ref{apx:naming}. $\rUNK$ indicates unknown results.}
\end{figure}

\paragraph{Terminating Linear programs}
The results of the experiments on these programs are shown in Table~\ref{fig:termlin}. Since \tool\ is nondeterministic, we ran our experiments 5 times.
We depict the ranking functions learned by \tool\ in the second column, taken from the first iteration of \tool. If a benchmark program
has more than one loop, we report the ranking functions learned from the last analyzed loop.
We also break down the overall time (and result) of \tool\ into time spent to learn ranking functions versus validate them.
Finally, we report the {\bf Total} time, averaged over 5 runs, as well as the standard deviation $\sigma_5$.
%

% \begin{wrapfigure}[17]{r}[40pt]{0.6\textwidth}
\begin{figure}
  \includegraphics[width=0.45\textwidth]{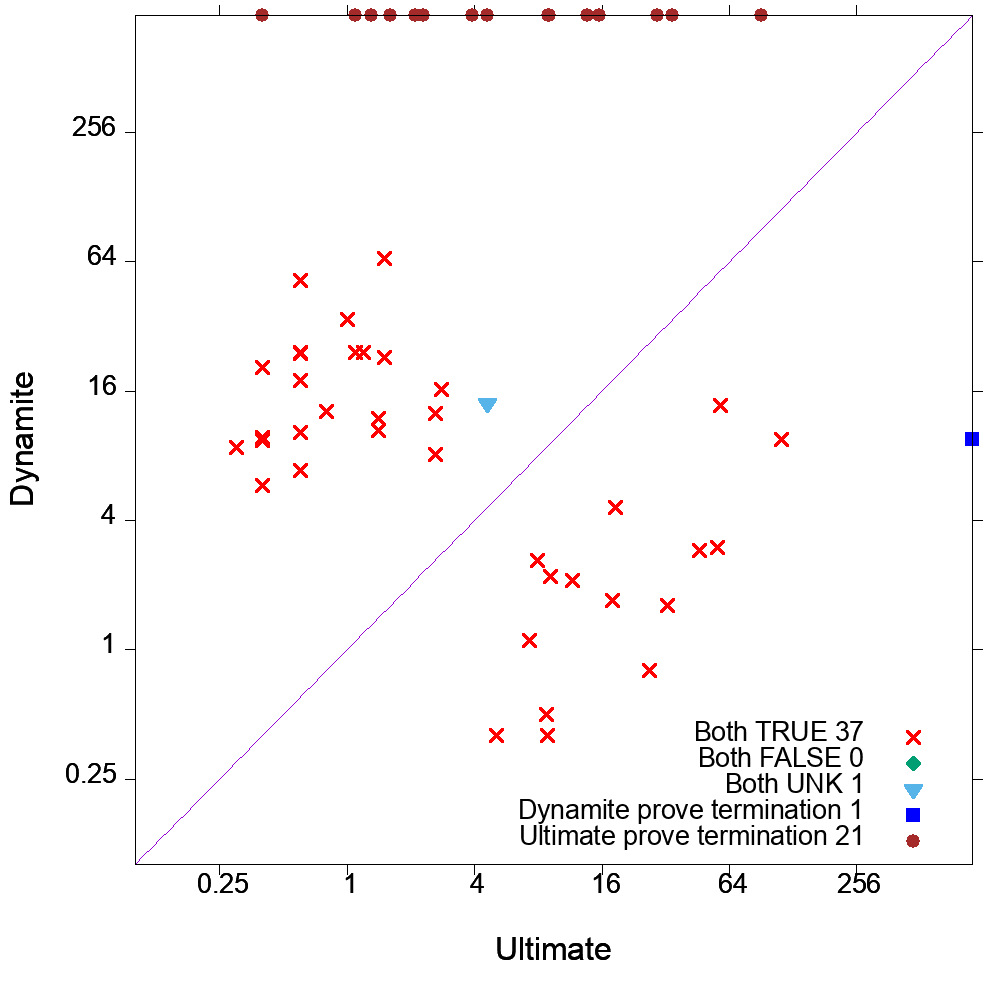}
  \caption{\label{fig:termlinplot} Visual comparison between \tool\ and \UAutomizer\ on linear termination benchmarks.}
% \end{wrapfigure}
\end{figure}
In the last two columns, we show the time and result of \UAutomizer.
These results are also visualized in the plot in Fig. \ref{fig:termlinplot}.
The results show that \UAutomizer\ often performs much faster than \tool\ on these linear examples, owing largely to the fact that \tool\ must execute the program many times (on the newly generated inputs), as is typically the case for data-driven strategies~\cite{nguyen2017counterexample}.
Nonetheless, the results show that \tool\ is competitive. 
In most benchmark programs, \tool\ can learn useful ranking functions from their terminating traces. For ranking functions that cannot be validated by the reachability provers, we manually checked if they are valid with respect to the observed terminating traces. We found that some of them are the desired ranking functions to prove the programs' termination while the others are still in good progress so that we could infer the desired ranking functions from their validation's counterexamples.
There are no ranking functions learned from unsupported recursive programs and string-manipulating programs.
We also found that \tool\ was able to infer a ranking function for the program \textts{HeHoLePo-ATVA2013-Fig5.c}, which can be validated successfully, while \UAutomizer\ cannot.
It is worth noting that \UAutomizer\ is a very mature tool, with contributions from multiple researchers/developers, has been applied in industrial settings, and has consistently performed well in the SV-COMP Termination categories. By contrast, \tool\ is still in its infancy. Furthermore, we will soon discuss non-linear programs, a class of programs not currently supported by \UAutomizer.

\paragraph{Non-Terminating Linear programs}

Of the \textts{termination-crafted-lit} SV-COMP suite, only 5 benchmarks were for non-termination. We ran \tool\ and Ultimate on them; the results are given in Fig.~\ref{fig:termlin}. Again we report the mean and standard deviation over 5 runs. In all cases, \UAutomizer\ was able to generate a lasso counterexample to termination.
\tool\ was able to produce a validated recurrent set in \diff[4]{3} programs with \textts{UPPERBOUND=3} in \ProveNT. The program \diff[\textts{ChCoFuNiHe-TACAS2014-Intro.c} has a nondeterministic loop body while]{} \textts{HeJhMaSu-POPL2002-LockEx.c} has a nondeterministic loop condition so that \diff[the loop condition and]{} $True$\diff[, respectively, are their]{is its} trivially valid recurrent set. Therefore, there is no cost for learning recurrent sets in these programs. \diff{On the other hand, the program \textts{ChCoFuNiHe-TACAS2014-Intro.c} has a nondeterministic assignment in its loop body so that while the loop condition is a (closed) recurrent set, it cannot be validated without an underapproximation to restrict the choice of nondeterministic values in that assignment.} The two programs \textts{Ur-WST2013-Fig1.c} and \textts{Velroyen.c} have many branches in their loop bodies but only some of them were taken by the symbolic execution tool to build the loop summaries. Unfortunately, in \textts{Ur-WST2013-Fig1.c}, the non-terminating branch was missing so that \tool\ cannot find any valid recurrent set from the other (terminating) branches in the summary.

\begin{figure}\footnotesize
\def\arraystretch{0.9}
\figtitle{\tool\ and Ultimate on LIN Non-termination Programs from SV-COMP}
\begin{tabular}{|p{1.6in}||p{1.0in}|rc|rc|rr||rc|}
\hline
\TCLheader{Rec.~Sets}
\hline
BrMaSi-CAV2005-Fig1-mod.c & $ y1{\neq}y2 \wedge y2=0 $ & 11.7  & \rTRUE & 8.9  & \rTRUE &  44.3 &   1.7 &   \ULTIMATE{  0.1   & \rFALSE } \\
ChCoFuNiHe-TACAS2014-Intro.c & & 5.8  & \rUNK & 2.7  & \rUNK &  38.68 &   5.4 &   \ULTIMATE{  0.2   & \rFALSE } \\
HeJhMaSu-POPL2002-LockEx.c & $ True $ & 0.0  & \rTRUE & 0.1  & \rTRUE  &  17.4 &   0.1 &   \ULTIMATE{  0.1   & \rFALSE } \\
Ur-WST2013-Fig1.c          & $ $ & 6.8  & \rUNK & 0.8  & \rUNK  &    17.7  &   1.8 &   \ULTIMATE{  0.4   & \rFALSE } \\
Velroyen.c                & $ x{\neq}0 $ & 0.0  & \rTRUE & 0.1  & \rTRUE  &  12.1 &   0.7 &   \ULTIMATE{  2.2   & \rFALSE } \\
\hline
%BradleyMannaSip           & $ y2, y1          $ & 6.4  & \rTRUE & 375.7 & \rUNK  &  71.5 &   \ULTIMATE{  0.1   & \rFALSE } \\
%Velroyen.c                & $ x, -x           $ & 0.9  & \rTRUE & 5.6  & \rUNK  &   8.4 &   \ULTIMATE{  2.2   & \rFALSE } \\
%Urban-WST2013-F           & $ 10 + -x         $ & 0.0  & \rTRUE & 11.3 & \rUNK  &  14.1 &   \ULTIMATE{  0.4   & \rFALSE } \\
%Urban-WST2013-F2          & $ x2              $ & \rAF & \rAF   & \rAF & \rAF   &  \rTO &   \ULTIMATE{  0.4   & \rFALSE } \\
%HenzingerJhalaM           & $ -got_lock       $ & \rAF & \rAF   & \rAF & \rAF   & 395.7 &   \ULTIMATE{  0.1   & \rFALSE } \\
%ChenCookFuhsNim           & $ i, 1 + -i, 4 + -i  $ & 0.5  & \rTRUE & 6.0  & \rUNK  &   8.5 &   \ULTIMATE{  0.2   & \rFALSE } \\
%\hline
%\input{termcraftlitnt}
\end{tabular}
\caption{\label{fig:termlin} Results of applying ultimate and \tool\ on the 5 \emph{non}-termination benchmarks from SV-COMP \textts{termination-crafted-lit}. Ultimate returned ``incorrect,'' indicating that it had found a non-terminating lasso to disprove termination.}
\end{figure}

\subsection{Termination of NLA programs}
\label{subsec:nlat}

Currently, we lack \diff[available]{challenging} benchmark suites for termination of non-linear programs. \diff{The existing \textts{polyrank} benchmark~\cite{Bradley2005poly} has only one (quadratic) polynomial program. The other programs in \textts{polyrank} are linear and many of them were included into the SV-COMP \textts{termination-crafted-lit} suite. \tool\ can prove the termination of 8/11 benchmarks (see Fig. \ref{fig:polyrank}) by inferring multiple linear ranking functions, instead of a single non-linear ranking function, and successfully validating them with Ultimate or CPAchecker. For the remaining 3 examples, DynamiTe was able to infer the correct ranking functions, but the validators could not validate them before timeout.} In order to better evaluate the tool, we adapted an existing non-linear testsuite from SV-COMP called \textts{nla-digbench} which consists of 28 programs implementing mathematical functions such as \textts{intdiv}, \textts{gcd}, \textts{lcm}, \textts{power}. Although these programs are relatively small (under 50 LoCs) they contain nontrivial structures such as nested loops and nonlinear invariant properties. To the best of our knowledge, \textts{nla-digbench} contains the largest number of programs containing nonlinear arithmetic. These programs have also been used to evaluate other numerical invariant systems~\cite{yao2020learning,rodriguez2007generating}.

However, these benchmarks are for invariant generation rather than termination \diff{and most of them are linear programs (with non-linear invariant properties)}. We therefore adapted these benchmarks to make them suitable non-linear examples for termination. For each benchmark, we manually examined the behavior of the program. The benchmarks contain commented nonlinear assertions, that illustrate the need for non-linear reasoning. For example, \textts{bresenham1} contains the assertion
\textts{2*Y*x - 2*X*y - X + 2*Y - v == 0}. We adapted these assertions to be loop conditions in various ways, creating one or more termination challenge programs. We typically geared the loop condition to the assertion. In this case, we used the invariant that the LHS is 0 and added an additional term \textts{+c} that increased on each iteration, and made the loop condition bounded by a variable \textts{k}. For example, from the \textts{bresenham1} program, we created a new program in which the loop condition is \textts{2*Y*x - 2*X*y - X + 2*Y - v + c <= k}.
In other cases, we introduced new variables and had them be integrated by other expressions that we knew to be monotonically increasing or replace variables and numbers with non-linear expressions that are equal to them.
We have made these {38} benchmarks available in the \materials{} and will issue a pull request to submit them to the SV-COMP benchmark repository. 

The results of applying \tool\ to these benchmarks is given in Fig.~\ref{fig:nlaterm}. For each benchmark, we give a brief description of the mathematical behaviors of the program in the second column. Based on the results, we display the output list of inferred ranking functions, as well as a breakdown of the time it took to learn versus validate them. As mentioned in the previous section, we use Ultimate and CPAchecker for validation. In {34} of the {38} benchmarks, \tool\ was able to guess ranking functions. The ranking functions derived from {8} of those {32} benchmarks can be validated. For those ranking functions that cannot be validated by the existing safety provers, we manually inspected them and confirmed they were correct.
In the remaining {4} cases, the program \textts{cohencu4} is non-terminating because the increment statement \textts{c++} was unintentionally not added. Therefore, there are no terminating traces to learn ranking functions. After fixing that problem, \tool\ can infer the desired ranking functions for this program successfully. The 2 programs \textts{freire1} and \textts{knuth-nosqrt} are originally floating-point programs but were intentionally transformed to integer programs. However, the invariant assertions in the original benchmarks is no longer valid in our adapted benchmark programs. Therefore, the desired ranking functions cannot be found from them. The last program \textts{knuth} still has the use of \textts{sqrt} function which is not supported by our CIL instrumentation.
It is worth noting that \UAutomizer\ cannot handle these benchmarks. In addition, since the validation get stuck on most benchmarks, we do not report the total time in Fig.~\ref{fig:nlaterm}.

\newcommand\NLAheader[1]{ &  & \multicolumn{1}{c}{\tool} & \multicolumn{2}{c}{{\bf Guessing}}  & \multicolumn{2}{c}{{\bf Validation}}  & \multicolumn{2}{c|}{\bf Total}\\%
{\bf Benchmark} & {\bf Desc.} & {\bf Learned #1} & Time & Res &  Time & Res &  Time  & Res \\}
\newcommand\NLAheaderNT[1]{ &  & \multicolumn{1}{c}{\tool} & \multicolumn{2}{c}{{\bf Learning}}  & \multicolumn{2}{c|}{{\bf Validation}}  \\%
{\bf Benchmark} & {\bf Desc.} & {\bf Learned #1} & Time & Res &  Time & Res  \\}
\newcommand\NLAheaderNTNoDesc[1]{ & \multicolumn{1}{c}{\tool} & \multicolumn{2}{c}{{\bf Learning}}  & \multicolumn{2}{c|}{{\bf Validation}}  \\%
  {\bf Benchmark} & {\bf Learned #1} & Time & Res &  Time & Res  \\}

\begin{figure}\footnotesize
  \def\arraystretch{0.9}
  \figtitle{\tool\ on NLA Termination Programs adapted from SV-COMP}
  \begin{tabular}{|l||p{2.5in}|rc|rc|}
    \hline
    \NLAheaderNTNoDesc{Ranking Functions}
    \hline
    \input{polyrank}
    \hline
  \end{tabular}
  \caption{\label{fig:polyrank} Results of applying \tool\ to the benchmark suite \textts{polyrank}.}
\end{figure}

\begin{figure}\footnotesize
\def\arraystretch{0.9}
\figtitle{\tool\ on NLA Termination Programs adapted from SV-COMP}
\begin{tabular}{|l|l||p{1.2in}|rc|rc|}
\hline
\NLAheaderNT{Ranking Fns.}
\hline
\input{nlaterm2}
\hline
\end{tabular}\\
\begin{tabular}{llp{9cm}}
Full RF for & \textsf{egcd3}:& $c + -v, v, -37\cdot b + -d, -293\cdot b + -d, -2341\cdot b + -d, -18725\cdot b + -d, -37449\cdot b + c + -d, 10083\cdot c + -d, 322639\cdot c + -d, 2581111\cdot c + -d, -1677722\cdot b + -d$\\
& \textsf{freire1}: & $-1\cdot r + 12\cdot a + 3\cdot k, -1\cdot r + 19\cdot a, -1\cdot r + 28\cdot a + 3\cdot k, -1\cdot r + 2\cdot k, -1\cdot r + -1\cdot a, -1\cdot r + -2\cdot a, -1\cdot r + -16\cdot a, -1\cdot r + 25\cdot k, -1\cdot r + 34\cdot k, -1\cdot r + 38\cdot k, -1\cdot r + 40\cdot k, -1\cdot r + 44\cdot k, -1\cdot r + -19\cdot a, -1\cdot r + 89\cdot k, -1\cdot r + 105\cdot k, -1\cdot r + 154\cdot k, -1\cdot r + 186\cdot a$
%&\textts{geo1}: & $c + -x, -c + x, -y, x + -y, y, k + -c     $ 
\end{tabular}
\caption{\label{fig:nlaterm} Results of applying \tool\ to our new benchmark suite of NLA termination challenge problems. For \textts{egcd3} and \textts{geo1} the full set of ranking functions are below the table. }
\end{figure}

\subsection{Non-Termination of NLA programs}
\label{subsec:nlant}

\diff{We first apply \tool\ to the existing non-linear non-termination benchmark {\sc Anant}~\cite{Cook2014}. The benchmark is a set of quadratic polynomial programs and some of them have non-determinism and divisions. The result of applying \tool\ on this benchmark is given in Fig. \ref{fig:anant}. \tool\ can prove the non-termination of 4 benchmarks that~\cite{Cook2014} cannot handle. However, there are 10 benchmarks that \tool\ cannot handle, due to non-determinism (4), overfitting invariants (1), overflow (1), and problems in SMT solvers (2) or in symbolic execution (2). This result shows that our dynamic approach is orthogonal to the existing static techniques for proving non-termination of (non-linear) programs.}

\diff[We]{In addition to the benchmark {\sc Anant}, we} also adapted SV-COMP \textts{nla-digbench} suite  to create NLA \emph{non}-termination challenge programs (\eg\ up-to-sextic polynomial programs). (These are also available in the supplementary materials and will be submitted to SV-COMP.) The results of applying \tool\ on this benchmark are given in Fig.~\ref{fig:nlanonterm}. Out of the {39} benchmarks, \tool\ was able to generate a recurrence set for {34} programs. 

Interestingly, we found that, for non-termination, our algorithm's semantic extraction of the first candidate recurrent set from the loop condition already provides a good guess to start with. Consequently, in 28 cases, dynamic analysis was actually unnecessary because our algorithm could already use the loop condition to guide guessing for a recurrent set. 
This is in contrast with termination, where we don't have have any semantic information to provide a starting guess for rank functions. 
%\red{wording of this par}
However, in 5 cases, dynamic refinement was necessary where the loop conditions are not the existing invariant assertions in the original benchmark programs. 
%In the remaining 6 cases, \tool\

%\chanh{what makes a program hard or easy?}
%The recurrent sets generated \chanh{are interesting because...}

\begin{figure}\footnotesize
  \def\arraystretch{0.9}
  \setlength{\tabcolsep}{0.2em}
  \figtitle{\tool\ on NLA Non-Termination Programs from {\sc Anant}}
  \begin{tabular}{|l|p{3in}|rc|rc|rc|}
    \hline
    \NLAheaderNTNoDesc{Recurrent Sets}
    \hline
    \input{anant}
    \hline
  \end{tabular}
  \caption{\label{fig:anant} Results of applying \tool\ to the benchmark suite {\sc Anant} of NLA non-termination problems.}
\end{figure}

\begin{figure}\footnotesize
\def\arraystretch{0.9}
\setlength{\tabcolsep}{0.2em}
\figtitle{\tool\ on NLA Non-Termination Programs adapted from SV-COMP}
\begin{tabular}{|l|l|p{2.5in}|rc|rc|rc|}
\hline
\NLAheaderNT{Rec. Sets}
\hline
\input{nladigbenchnt}
\hline
\end{tabular}
\caption{\label{fig:nlanonterm} Results of applying \tool\ to our new benchmark suite of NLA termination challenge problems.}
\end{figure}

\newcommand\gRS{rcr.}
\newcommand\gRF{rf.}
\newcommand\INTHEAD{ & &  & \multicolumn{4}{c|}{ \tool }\\%
{\bf Benchmark} &  {\bf \#L} & {\bf Exp.} &   Out & \#Sw. & Res &  Time  \\}
\begin{figure}\footnotesize\centering
\figtitle{\tool's \ProveTNT\ Algorithm on NLA Term.~\& Non-Term.~Programs adapted from SVCOMP}
\begin{tabular}{rcl}
\begin{minipage}{2.5in}\footnotesize\centering
\def\arraystretch{0.9}
\setlength{\tabcolsep}{0.3em}
\begin{tabular}{|l|c|c||c|r|r|r|r|r|}
\hline
\INTHEAD
\hline
\input{nlaintegrated3}
\hline
\end{tabular}
\end{minipage}
&\hfill&
\begin{minipage}{2.5in}\footnotesize\centering
\def\arraystretch{0.9}
\setlength{\tabcolsep}{0.3em}
\begin{tabular}{|l|c|c||c|r|r|r|r|r|}
\hline
\INTHEAD
\hline
\input{nlaintegrated3b}
\hline
\end{tabular}\\
$\;$ $\;$
\end{minipage}
\end{tabular}
\caption{\label{fig:nlaintegrated} Results of applying \tool\ to a mix of terminating and non-terminating examples. }
\end{figure}

\subsection{Integrated Algorithm: Discriminating between termination and non-termination}

We experimented with \ProveTNT\ to evaluate (a) whether the algorithm is able to discriminate programs that terminate from those that non-terminate and (b) whether feedback from a failed attempt to prove termination can inform a proof of non-termination and vice-versa. For (a), we jumbled together all of the NLA benchmarks and ran the integrated algorithm on them. The results are given in Fig.~\ref{fig:nlaintegrated}. For these benchmarks we note the number of loops (\#L). We also indicate whether a guess was made of either a recurrent set (rcr) or a ranking function (rf). \ProveTNT\ makes an initial guess whether to pursue termination or non-termination and if the choice fails, ``switches'' to the opposite tack. We report the number of switches (\#Sw), as well as the final validated conclusion and the total time. 

For the vast majority of the examples, there are no switches, which means the initial choice (based on dynamic execution sampling the instrumented program) was was a good one, or that \ProveTNT\ timed out before it could validate a guess. In 16 of the {77} benchmarks, a switch was made at least once. 
As compared with Fig.~\ref{fig:nlaterm} (NLA Termination) and Fig.~\ref{fig:nlanonterm} (NLA Non-termination), more timeouts occur here. However, the comparison is a little unfair: in the earlier experiments we already knew the conclusion (T versus NT) so we aimed \tool\ toward the prize. 

For \ProveTNT, one pitfall is that a wrong initial choice could lead to time spent attempting to  validate a ranking function, when it should be spent pursing recurrent sets (or vice-versa). 
A first attack against this problem is to improve the first guess: the better the initial guess, the closer the results come to those in Fig.~\ref{fig:nlaterm} and Fig.~\ref{fig:nlanonterm}.
A na\"{i}ve strategy could be to add a timeout to validation. Another could be to parallelize, pursuing  termination and non-termination concurrently. The downside of parallelization, is that one cannot use the output of one failed endeavor to inform the other. 
\ProveTNT\ takes an alternate strategy, as dicussed in Section~\ref{sec:tnt}: pursue one and, if it fails, exploit the information from the counter example to expedite the alternative. In the worst case, this at least improves over running the two strategies sequentially. In Section~\ref{sec:tnt} we gave two examples that demonstrate where the integrated strategy helps.

\subsection{Discussion}
Unlike static analysis techniques, our dynamic analysis technique executes the programs to collect data in the form of snapshots at several program locations. In some cases, the time to execute the programs and process their raw output is significant, especially on programs with high-complexity or programs with a large number of parameters which require a large number of random inputs to maintain the data's diversity. On the other hand, when there is not enough data, overfitting may occur. In proving non-termination, overfitting can make the dynamically inferred conditions too strong to refine a recurrent set. In proving termination, overfitting may create a large number of ranking functions and overwhelm the validation tools. We also have a problem with branching in the loop body where the loop summary returned by the symbolic execution is imprecise since some branches are not taken. That imprecision affects the refinement of  candidate recurrent sets.

On the other hand, our dynamic analysis has some advantages that static analysis does not have. For example, we can find a reliable set of ranking functions from known terminating traces at the beginning so we can avoid many expensive validation steps whereas static analysis techniques require many of them to refine the ranking function set from scratch. 

\section{Related Work}
\label{sec:related}

%Recent years have seen a variety of techniques that  look beyond linear to \emph{non}-linear arithmetic domains and/or exploit dynamic analysis to bolster verification for safety, termination and non-termination. We now summarize some of these works.
%
%In this work, we proposed an algorithm that 
%incorporates learning-based reasoning for guessing invariants
%(for nontermination) and rank functions (for termination)
%into an integrated algorithm that samples inputs and
%partitions them into  terminating versus potentially
%non-terminating traces.  We then perform learning on these 
%partitions separately, and combine counterexamples from one partition
%to inform the other.

%
%To our knowledge, no existing works
%uniform idea for both t and nt.
%lift dynamic invariant generation so that it can be used 
%for guessing RS and RF
%   RF: only generate traces/states
%   use DIG for invariant inside loop body as candidate RS.

\paragraph{Inference of non-linear invariants.}
Nonlinear polynomial relations arise in many safety-and security-critical applications.
For example, the Astr\'ee analyzer, which has been applied to verify the absence of errors in the Airbus A340/A380 avionic systems~\cite{blanchet2003static}, implements the ellipsoid abstract domain~\cite{feret2004static} to represent and analyze a class of quadratic inequalities.

\citet{rodriguez2007generating,rodriguez2007automatic} used abstract interpretation to infer nonlinear equalities.
They first observe that a set of polynomial invariants form the algebraic structure of an ideal, then compute the polynomial invariants using Grobner basis and operations over ideals based on the structure of the program until a fixed point is reached.
%Scalability is an issue because the approach must consider all possible program paths to guarantee sound results.
The approach is restricted to non-nested loops and programs with assignments and loop guards expressible as polynomial equalities.
The SPEED project~\cite{gulwani2009speed, gulwani2009speedb} uses a numerical abstract domain~\cite{gulavani2008numerical} to compute disjunctive and non-linear invariants representing  runtime complexity bounds.
The numerical domain uses operators such as \texttt{max} to represent disjunction and constraints over various operators using inference rules to represent nonlinear operators.

%The approach is also restricted to non-nested loops.

%Sharma et al. proposed a hybridizes dynamic and static analysis to generate equality invariants. They use DIG's technique to compute nonlinear equality invariants from traces and use SMT solving to verify that the candidate invariants are correct with respect to the program source code. Counterexamples to candidate invariants that fail to prove are subsequently used to produce more traces to generate better candidate invariants.
The well-known dynamic invariant tool Daikon~\cite{ernst2007daikon, ernst2001dynamically} infers candidate invariants under various templates over concrete program states. 
The tool comes with a large set of templates which it tests against observed traces, removing those that fail, and return the remaining ones as candidate invariants.
DIG~\cite{tosem2013}, which is used by \tool\, focuses on numerical invariants and therefore can compute more expressive (e.g., nonlinear) numerical relations than those supported by Daikon's templates.
%The pure dynamic analysis in DIG supports numerical invariants of the forms nonlinear equalities~\cite{nguyen2012using}, octagonal inequalities~\cite{tosem2013}, and max/min-plus inequalities~\cite{nguyen2014using}.

%Dynamic analysis also has been applied to to infer invariants in separation logic to reason about memory properties~\cite{pldi2019}.
More recently, \citet{yao2020learning} described a method for inferring invariants through a form of neural networks.
The technique uses a Continous Logic Network to learn SMT formulas directly from program traces. The authors show that this approach can learn more general nonlinear invariants  (equalities, inequalities, and disjunction). % than those from~\cite{numinv}.

There are several hybrid works in the form of guessing and checking invariants.
In \citet{sharma2013data}, ``guess'' component infers nonlinear equalities using the similar equation solving technique in DIG and the ``check'' component uses the Z3 SMT solver.
Counterexamples from the checker are used to produce more traces to infer better invariants.
%This work extends DIG's algorithms to infer nonlinear equalities as well as linear iequalities and max- and min-plus invariants and verifies them using a customized k-inductive prover.
The works from NumInv~\cite{nguyen2017counterexample} and SymInfer~\cite{nguyen2017symlnfer} combined the dynamic analyis from DIG to infer nonlinear invariants with symbolic execution to remove spurious results. %The obtained results are only sound up to the depth used in symbolic execution.

PIE~\cite{padhi2016data} and ICE~\cite{garg2014ice} also use an guess and check approach to infer invariants to prove a given specification. To prove a property, PIE iteratively infers and refined invariants by constructing necessary predicates to separate (good) states satisfying the property and (bad) states violating that property. ICE uses a decision learning algorithm to guess inductive invariants over predicates sepa- rating good and bad states. The checker produces good, bad, and ``implication'' counterexamples to help learn more precise invariants.
%For efficiency, they focus on octagonal predicates and only search for invariants that are boolean combinations of octagonal relations. %In general, these techniques focus on invariants that are necessary to prove a given specification and, thus, the quality of the invariants are dependent target specification.

%% One (semi-automatic) technique to compute such disjunctive and non-linear invariants is to use the numerical abstract domain described in [6]. The numeri- cal abstract domain is parametrized by a base linear arithmetic abstract domain and is constructed by means of two domain lifting operations that extend the
%% base linear arithmetic domain to reason about the max operator and other op- erators whose semantics is specified using a set of inference rules. One of the domain lifting operation extends the linear arithmetic domain to represent lin- ear relationships over variables as well as max-expressions (an expression of the form Max(e1, . . . , en) where ei’s are linear expressions). Another domain lifting operation lifts the abstract domain to represent constraints not only over pro- gram variables, but also over expressions from a given finite set of expressions S. The semantics of the operators (such as multiplication, logarithm, etc.) used in constructing expressions in S is specified as a set of inference rules. The ab- stract domain retains efficiency by treating these expressions just like any other variable, while relying on the inference rules to achieve precision.

\paragraph{Termination}
Today, numerous theories, techniques, and tools exist for proving termination and non-termination~\cite{Podelski2004,Giesl2004,Cook2006,Cook2011,Cousot2012,Dietsch2015}. 
Tools include Terminator~\cite{Cook2006}, Ultimate Automizer~\cite{ultimate}, {\sc HipTNT+}~\cite{Le2015}, \citet{function}, \citet{cpachecker}, and \citet{aprove}.
There is even a category on termination in the Software Verification Competition (SV-COMP)~\cite{svcomp2020}.
Along the way, some have shown methods for \emph{conditional} termination, whereby preconditions are found that specify the portion of traces that terminate~\cite{Cook2008,Le2015}.
Another active line of research has focused on flavors of ranking functions,
including piece-wise~\cite{Urban2013},
ordinals~\cite{Urban2014},
size-change~\cite{Lee2001}, and
lexicographic~\cite{Bradley2005}.
\citet{Babic2007} focused on proving termination of a restricted class of non-linear loops, called NAW loops, which have special properties to allow their termination to be proved via analyzing the divergence of variables influencing the loop conditions. 

%
%\citet{Babic2007} focused on proving termination of a restricted class of non-linear loops, called NAW loops, which have special properties to allow their termination to be proved via analyzing the divergence of variables influencing the loop conditions. Along the line of research on proving non-termination, \citet{Gupta2008} introduced a constraint solving technique to find recurrent sets of non-terminating loops. \citet{Chen2014} strengthened the concept of recurrent sets so that they can reduce the non-termination problem to safety proving and support more nondeterministic programs. There are some other approaches that attempt to reason program termination and non-termination at the same time. \citet{Harris2010} introduced a static technique that maintains an over- and under-approximation for alternatively proving termination or non-termination of a program. \citet{Le2014} proposed a resource logic which can uniformly specify and verify preconditions of program termination and non-termination. Later, in \cite{Le2015}, they introduced a second-order constraint-based technique to derive termination summary in the form of that logic automatically.

\diff{\citet{Bradley2005poly,Bradley2005} focused on the class of polynomial loops from which finite different trees can be derived. However, the techniques could not work on examples with infinite difference trees.}
\begin{wrapfigure}[3]{r}[30pt]{0.35\textwidth}
  \vspace*{-1em}
  \begin{lstlisting}[language=Python,style=pnnstyle]
    if x >= 0:
      while x * x <= 100:
        x = 2 * x + 1
  \end{lstlisting}
\end{wrapfigure}
\diff{For example, to prove the termination of the program on the right, those techniques construct a difference tree whose root is the expression \textts{100 - x * x} in the loop condition. Since the tree is infinite, they could not prove the program's termination. \tool\ can derive the ranking function \textts{10 - x} from concrete snapshots of that example, which is sufficient to prove its termination.}

A number of works have exploited dynamic information to inform termination reasoning. 
%Termination proofs from tests. 
\citet{Nori2013} showed that linear regression can be used to \diff{dynamically} infer bounds of program loops \diff{from test suites} and these bounds imply termination. They then attempt to validate those bounds and use counterexamples to improve the precision of inference. \diff{By using the disjunctive well-foundedness in the termination proofs, \tool\ can prove the termination of examples in~\cite{Nori2013} which have a disjunctive or non-linear bound with only simple linear ranking functions.}
\citet{Nguyen2019} describe runtime contracts for enforcing termination, using the size-change strategy for termination.

Several static techniques are able to infer polynomial \emph{resource bounds}~\cite{Hoffmann2010a,Hoffmann2010b,Hoffmann2011}.
The TiML functional language~\cite{wang2017timl} allows a user to specify time complexity as types and then uses type checking to verify the specified complexity.
The WISE tool~\cite{burnim2009wise} uses concolic execution to search for a path policy that leads to an execution path with high resource usage.

\paragraph{Non-Termination} Along the line of research on proving non-termination, \citet{Gupta2008} introduced a constraint solving technique to find recurrent sets of non-terminating loops. Later, \citet{Chen2014} strengthened the concept of recurrent sets \diff{to ``closed'' recurrent sets} so that they can reduce the non-termination problem to safety proving and support more nondeterministic programs.

\diff{\citet{Cook2014} proved non-termination of non-linear programs by soundly over-approximating the programs to nondeterministic linear programs and then using~\citet{Chen2014} approach to disprove their termination. However, since the technique searches for linear recurrent sets via Farkas' lemma on the abstract linear programs, it cannot generate recurrent sets described by non-linear equations. For example, in the benchmarks from Figure \ref{fig:nlanonterm}, there were only 5 cases where \tool\ learned a linear recurrent set and in roughly half of the cases, \tool\ learned a non-linear recurrent set, which could not be found using the~\citet{Cook2014} approach. Therefore, while we are able to leverage ongoing advances in non-linear invariant generation techniques (a growing area of research), the~\citet{Cook2014} approach cannot. In addition,~\citet{Cook2014} build over-approximation by using an abstract interpreter, such as Interproc, which usually does not perform well on non-linear programs. As shown in Section~\ref{subsec:nlant}, \tool\ can prove the non-termination of all 4 {\sc Anant} benchmarks in~\cite{Cook2014} that they cannot handle.}

\diff{\citet{Frohn2019} utilized recurrence relation solvers to replace loops whose non-termination cannot be proved by loop-free transitions in finding feasible paths to a non-terminating loop. The technique relies on recurrence relation solvers, whose supporting forms of recurrence relations are restricted. For example, the approach cannot prove the non-termination of the \textts{p3} program in the aforementioned non-linear {\sc Anant} benchmarks while \tool\ can.}

There are some other approaches that attempt to reason program termination and non-termination at the same time.
\citet{Harris2010} introduced a technique that maintains an over- and under-approximation for alternatively proving termination or non-termination of a program.
\citet{Le2014} proposed a resource logic which can uniformly specify and verify preconditions of program termination and non-termination.
Later, \citet{Le2015} introduced a second-order constraint-based technique to derive termination summary in the form of that logic automatically. \diff{However, they cannot handle non-linear programs.}

%For dynamic techniques, Goldsmith et al. [35] propose a technique to measure empirical complexity by running the program on work- loads of various sizes and fit the program runtime into a model that predicts performance. PerfSyn [22] uses a black-box evolutionary search that mutates program code to identify performance bottlenecks. Similarly, Slowfuzz [70] (Section 4.3) is a blackbox fuzzer that mutates bytecodes to identify high-complexity inputs. Sigularity [81] uses genetic programming to evolve programs in domain-specific languages to represent high-complexity input patterns.

%% \paragraph{Guess and Check Method} Recently, several techniques (such as PIE , ICE, iDiscovery, NumInv) have been developed to infer numerical invariants using a hybrid approach that dynamically infers candidate invariants and then statically checks them against the program code. 
%% Instead of repatedly invoking a static analysis or tool, the SymInfer tool improves the guess-and-check approach by extracts symbolic states from a symbolic execution tool once and re-use these symbolic states to verify and iteratively refine the invariant generation process.

\section{Conclusion}

We have shown that dynamic strategies for discovering invariants and sampling transitive closure can be incorporated with static refinement into an overall framework for proving termination or non-termination of nonlinear programs. \citet{github} is publicly available and the new benchmark suites \textts{nla-term} and \textts{nla-nonterm} will soon be submitted to SV-COMP. While \tool\ already exploits concurrency by simultaneously attempting validation with CPAchecker and Ultimate, as well as within DIG, one avenue for improvement is to parallelize \ProveTNT.

\section*{Acknowledgment}
We thank the anonymous reviewers for the helpful feedback.
Ton Chanh Le, Timos Antonopoulos, and Eric Koskinen are supported by the Office of Naval Research under Grant N00014-17-1-2787.
ThanhVu Nguyen is supported by the National Science Foundation under Grant CCF-1948536 and the Army Research Office under Grant W911NF-19-1-0054.

%
%
%%% Acknowledgments
%\begin{acks}                            %% acks environment is optional
%                                        %% contents suppressed with 'anonymous'
%  %% Commands \grantsponsor{<sponsorID>}{<name>}{<url>} and
%  %% \grantnum[<url>]{<sponsorID>}{<number>} should be used to
%  %% acknowledge financial support and will be used by metadata
%  %% extraction tools.
%  This material is based upon work supported by the
%  \grantsponsor{GS100000001}{National Science
%    Foundation}{http://dx.doi.org/10.13039/100000001} under Grant
%  No.~\grantnum{GS100000001}{nnnnnnn} and Grant
%  No.~\grantnum{GS100000001}{mmmmmmm}.  Any opinions, findings, and
%  conclusions or recommendations expressed in this material are those
%  of the author and do not necessarily reflect the views of the
%  National Science Foundation.
%\end{acks}

%% Bibliography
\bibliography{ejk}

%
%%%% Appendix
\vfill
\pagebreak
\appendix
\section{Full version of Fig.~\ref{fig:termlin}}
\label{apx:tcl}

\begin{center}\scriptsize
\def\arraystretch{0.9}
\figtitle{\tool\ and Ultimate on LIN Termination Programs from SVCOMP}
\begin{tabular}{|p{1.5in}||p{1.3in}|rc|rc|rr||rc|}
\hline
\TCLheader{Rank Functions}
\hline
AlDaFeGo-SAS2010-Fig1.c             & $ x, y            $ & 18.8 & \rTRUE & 24.9 & \rTRUE & 46.24 &   1.7 &   \ULTIMATE{  2.9   & \rTRUE } \\
AlDaFeGo-SAS2010-cousot9.c          & $ j, i            $ & 8.8  & \rTRUE & 6.1  & \rTRUE & 24.24 &   5.0 &   \ULTIMATE{  1.2   & \rTRUE } \\
AlDaFeGo-SAS2010-easy1.c            & $ -x, -x + 14\cdot z, -x + 18\cdot z, \etal$ & 42.9 & \rTRUE & 72.8 & \rUNK  & 112.34 &  38.7 &   \ULTIMATE{  0.4   & \rTRUE } \\
AlDaFeGo-SAS2010-easy2-2.c          & $ z               $ & 7.4  & \rTRUE & 8.0  & \rTRUE &  20.7 &   1.2 &   \ULTIMATE{  0.4   & \rTRUE } \\
AlDaFeGo-SAS2010-loops.c            & $ x               $ & 27.5 & \rTRUE & 18.4 & \rTRUE & 57.92 &   5.1 &   \ULTIMATE{  13.7  & \rTRUE } \\
AlDaFeGo-SAS2010-nestedLoop-1.c     & $ j, -k + n, k, -i + N  $ & 101.7 & \rTRUE & 54.0 & \rUNK  & 241.1 &  15.1 &   \ULTIMATE{  15.6  & \rTRUE } \\
AlDaFeGo-SAS2010-random1d-2.c       & $ max + -x        $ & 12.3 & \rTRUE & 6.6  & \rTRUE & 23.94 &   1.9 &   \ULTIMATE{  0.6   & \rTRUE } \\
AlDaFeGo-SAS2010-random2d.c         & $ N + r, -r, -i + r, -r + x, i + -r, N + -i  $ & 23.8 & \rTRUE & 13.2 & \rTRUE & 52.38 &   7.9 &   \ULTIMATE{  0.6   & \rTRUE } \\
AlDaFeGo-SAS2010-speedpldi2.c       & $ -v2, v2, v1     $ & 29.4 & \rTRUE & 5.3  & \rUNK  &  43.3 &   9.2 &   \ULTIMATE{  1.6   & \rTRUE } \\
AlDaFeGo-SAS2010-speedpldi3.c       & $ -j + n, j, -i + n  $ & 13.1 & \rTRUE & 15.0 & \rTRUE & 23.04 &   7.6 &   \ULTIMATE{  1.5   & \rTRUE } \\
AlDaFeGo-SAS2010-speedpldi4.c       & $ i               $ & 4.7  & \rTRUE & 2.7  & \rTRUE & 10.46 &   1.2 &   \ULTIMATE{  1.4   & \rTRUE } \\
AlDaFeGo-SAS2010-wcet2.c            & $ -i, 1 + -i, j, -i + j, -j, 2 + -i  $ & 11.5 & \rTRUE & 121.5 & \rUNK  & 148.14 &  75.3 &   \ULTIMATE{  1.3   & \rTRUE } \\
AlDaFeGo-SAS2010-while2.c           & $ i               $ & 17.0 & \rTRUE & 19.6 & \rTRUE & 34.48 &   7.4 &   \ULTIMATE{  1.0   & \rTRUE } \\
AlDaFeGo-SAS2010-wise.c             & $ x + -y, -x + y  $ & 8.4  & \rTRUE & 4.7  & \rTRUE & 17.88 &   2.2 &   \ULTIMATE{  1.7   & \rTRUE } \\
Av-FLOPS2006-Table1.c               & $ -i + y          $ & 14.2 & \rTRUE & 9.1  & \rTRUE &  26.8 &   2.0 &   \ULTIMATE{  0.8   & \rTRUE } \\
BrMaSi-CAV2005-Fig1.c               & $ y1, y2          $ & 6.3  & \rTRUE & 2.9  & \rTRUE &  12.6 &   1.0 &   \ULTIMATE{  2.6   & \rTRUE } \\
BrCoFu-CAV2013-Fig1.c               & $ -j + n, j, -i, -i + n  $ & 28.9 & \rTRUE & 14.2 & \rTRUE & 66.22 &  22.0 &   \ULTIMATE{  1.5   & \rTRUE } \\
BrCoFu-CAV2013-Fig9a.c              & $ -j + n, -i + n  $ & 29.9 & \rTRUE & 19.5 & \rTRUE & 56.38 &   5.3 &   \ULTIMATE{  3.0   & \rTRUE } \\
BrCoFu-CAV2013-Intro.c              & $ x               $ & 2.4  & \rTRUE & 3.1  & \rTRUE &  7.96 &   0.2 &   \ULTIMATE{  2.6   & \rTRUE } \\
ChCoGuSaYa-ESOP2008-easy1.c         & $ -x + -z, -x + 3\cdot z, \etal $ & 12.2 & \rTRUE & 32.6 & \rUNK  &  80.6 &  45.8 &   \ULTIMATE{  0.4   & \rTRUE } \\
ChCoGuSaYa-ESOP2008-easy2.c         & $ z               $ & 3.9  & \rTRUE & 2.8  & \rTRUE &   9.4 &   0.8 &   \ULTIMATE{  0.4   & \rTRUE } \\
ChCoGuSaYa-ESOP2008-random1d.c      & $ -a, max + -x    $ & 12.8 & \rTRUE & 4.0  & \rTRUE & 18.02 &   2.2 &   \ULTIMATE{  0.6   & \rTRUE } \\
ChCoGuSaYa-ESOP2008-random2d.c      & $ r, -r, -i + N   $ & 16.0 & \rTRUE & 6.3  & \rTRUE & 24.12 &   1.4 &   \ULTIMATE{  0.6   & \rTRUE } \\
ChFlMu-SAS2012-Ex2.20.c             & $ x               $ & 0.0  & \rTRUE & 3.0  & \rTRUE &  5.82 &   0.6 &   \ULTIMATE{  0.4   & \rTRUE } \\
ChFlMu-SAS2012-Ex3.01.c             & $ -x + z, y, -x   $ & 5.9  & \rTRUE & 4.0  & \rTRUE & 16.34 &   2.4 &   \ULTIMATE{  2.8   & \rTRUE } \\
CoSeZu-TACAS2013-Fig1.c             & $ x               $ & 1.7  & \rTRUE & 10.7 & \rUNK  &  10.5 &   2.9 &   \ULTIMATE{  1.3   & \rTRUE } \\
CoSeZu-TACAS2013-Fig7a.c            & $ d               $ & 0.3  & \rTRUE & 4.3  & \rUNK  &  7.32 &   0.2 &   \ULTIMATE{  2.1   & \rTRUE } \\
CoSeZu-TACAS2013-Fig7b.c            & $ z, x, y         $ & 2.6  & \rTRUE & 4.0  & \rTRUE &  9.18 &   0.5 &   \ULTIMATE{  2.2   & \rTRUE } \\
CoSeZu-TACAS2013-Fig8a-mod.c        & $ -x, x, -K + x, K + -x  $ & 11.7 & \rTRUE & 4.0  & \rTRUE & 18.62 &   0.9 &   \ULTIMATE{  4.6   & \rTRUE } \\
CoSeZu-TACAS2013-Fig8a.c            & $ x, -x           $ & 5.4  & \rTRUE & 5.2  & \rTRUE &  11.9 &   0.5 &   \ULTIMATE{  1.4   & \rTRUE } \\
CoSeZu-TACAS2013-Fig8b.c            & $ M + -x, x       $ & 7.4  & \rTRUE & 5.5  & \rUNK  &  14.7 &   2.5 &   \ULTIMATE{  4.6   & \rTRUE } \\
GoRe-CAV2006-Fig1a.c                & $ -x + y, y, 5 + -y, 20 + -y  $ & 0.2  & \rTRUE & 74.1 & \rUNK  & 100.5 &  76.5 &   \ULTIMATE{  1.1   & \rTRUE } \\
HaLaNoRa-SAS2010-Fig1.c             & $                 $ & \rUNK  & \rUNK  & \rUNK  & \rUNK  &  8.94 &   1.8 &   \ULTIMATE{  29.4  & \rTRUE } \\
HaLaNoRa-SAS2010-Fig3.c             & $                 $ & \rUNK & \rUNK   & \rUNK & \rUNK   &  1.94 &   0.0 &   \ULTIMATE{  0.4   & \rTRUE } \\
HeHoLePo-ATVA2013-Fig1.c            & $ x               $ & 1.6  & \rTRUE & 4.4  & \rTRUE &  8.14 &   0.2 &   \ULTIMATE{  2.6   & \rTRUE } \\
HeHoLePo-ATVA2013-Fig4.c            & $ x               $ & 1.5  & \rTRUE & 2.7  & \rTRUE &  6.82 &   0.2 &   \ULTIMATE{  0.6   & \rTRUE } \\
HeHoLePo-ATVA2013-Fig5.c            & $ x               $ & 2.3  & \rTRUE & 5.8  & \rTRUE &  9.56 &   0.5 &   \ULTIMATE{  3.8   & \rUNK  } \\
HeHoLePo-ATVA2013-Fig6.c            & $ y, x            $ & 0.3  & \rTRUE & 2.5  & \rTRUE &  5.06 &   0.1 &   \ULTIMATE{  0.4   & \rTRUE } \\
%HeJhMaSu-POPL2002-LockEx.c          & $ -got_lock       $ & 0.0  & \rTRUE & 4.5  & \rUNK  &  6.74 &   0.1 &   \ULTIMATE{  0.1   & \rFALSE } \\
KrShTsWi-CAV2010-Ex.c               & $ 77 + -i, 144 + -i, 228 + -i, 251 + -i, 252 + -i  $ & 3.6  & \rTRUE & 6.7  & \rUNK  & 13.54 &   0.7 &   \ULTIMATE{  0.4   & \rTRUE } \\
KrShTsWi-CAV2010-Fig1.c             & $ 28 + -x, 82 + -x, 88 + -x, \etal $ & 8.9  & \rTRUE & 6.0  & \rUNK  & 14.66 &   1.5 &   \ULTIMATE{  3.9   & \rTRUE } \\
LaOlRo-CaRu-FMCAD2013-Fig1.c        & $ y, x, z         $ & 8.7  & \rTRUE & 132.2 & \rTRUE & 112.42 &  88.8 &   \ULTIMATE{  9.5   & \rTRUE } \\
LeJoBe-Amram-POPL2001-Ex4.c         & $                 $ & \rAF & \rAF  & \rAF & \rAF  &  3.56 &   1.2 &   \ULTIMATE{  34.5  & \rTRUE } \\
LeJoBe-Amram-POPL2001-Ex5.c         & $                 $ & \rAF & \rAF  & \rAF & \rAF  &   4.6 &   0.2 &   \ULTIMATE{  14.0  & \rUNK  } \\
LeHe-TACAS2014-Ex1.c                & $ q               $ & 2.2  & \rTRUE & 2.9  & \rTRUE &  8.78 &   1.8 &   \ULTIMATE{  0.5   & \rTRUE } \\
LeHe-TACAS2014-Ex9.c                & $ p, q            $ & 2.2  & \rTRUE & 8.2  & \rTRUE & 12.92 &   3.4 &   \ULTIMATE{  0.8   & \rTRUE } \\
LeHe-WST2014-Ex9.c                  & $ x               $ & 2.0  & \rTRUE & 3.6  & \rTRUE &  8.88 &   2.5 &   \ULTIMATE{  0.4   & \rTRUE } \\
PoRy-LICS2004-Fig1.c                & $ y, x            $ & 8.0  & \rTRUE & 11.2 & \rUNK  & 16.84 &   3.9 &   \ULTIMATE{  13.6  & \rTRUE } \\
PoRy-TACAS2011-Fig1.c               & $ y               $ & 2.4  & \rTRUE & 4.6  & \rTRUE &  8.76 &   0.9 &   \ULTIMATE{  0.3   & \rTRUE } \\
PoRy-TACAS2011-Fig2.c               & $ y, x            $ & 7.0  & \rTRUE & 8.5  & \rTRUE & 24.14 &   0.9 &   \ULTIMATE{  1.1   & \rTRUE } \\
PoRy-TACAS2011-Fig4.c               & $ y, x            $ & 0.3  & \rTRUE & 5.3  & \rTRUE &  7.26 &   0.4 &   \ULTIMATE{  1.1   & \rTRUE } \\
%Ur-WST2013-Fig1.c                   & $ 9 + -x, 10 + -x  $ & 0.0  & \rTRUE & 7.7  & \rUNK  &    10 &   1.8 &   \ULTIMATE{  0.4   & \rFALSE } \\
Ur-WST2013-Fig2-mod1000.c           & $                 $ & 0.0  & \rUNK  & 3.9  & \rUNK  &  7.16 &   0.2 &   \ULTIMATE{  0.4   & \rFALSE } \\
Ur-WST2013-Fig2.c                   & $ -x1, -x1 + 5\cdot x2, \etal   $ & 14.2 & \rTRUE & 12.5 & \rTRUE & 32.78 &   5.7 &   \ULTIMATE{  1.6   & \rTRUE } \\
UrMi-ESOP2014-Fig3.c                & $ y, -x, x        $ & 3.1  & \rTRUE & 5.6  & \rTRUE & 11.58 &   0.8 &   \ULTIMATE{  2.1   & \rTRUE } \\
aviad.c                             & $ a               $ & 2.7  & \rTRUE & 5.4  & \rTRUE &  10.3 &   0.9 &   \ULTIMATE{  0.6   & \rTRUE } \\
cstrcmp.c                           & $                 $ & \rAF & \rAF   & \rAF & \rAF   &   1.9 &   0.1 &   \ULTIMATE{  2.3   & \rTRUE } \\
cstrcspn.c                          & $                 $ & \rAF & \rAF   & \rAF & \rAF   &     2 &   0.2 &   \ULTIMATE{  90.3  & \rTRUE } \\
cstrlen.c                           & $                 $ & \rAF & \rAF   & \rAF & \rAF   &   2.1 &   0.2 &   \ULTIMATE{  1.1   & \rTRUE } \\
cstrpbrk.c                          & $                 $ & \rAF & \rAF   & \rAF & \rAF   &  1.84 &   0.0 &   \ULTIMATE{  8.9   & \rTRUE } \\
cstrspn.c                           & $                 $ & \rAF & \rAF   & \rAF & \rAF   &  1.96 &   0.2 &   \ULTIMATE{  9.0   & \rTRUE } \\
genady.c                            & $ i               $ & 0.1  & \rTRUE & 7.4  & \rTRUE &  9.78 &   1.6 &   \ULTIMATE{  0.4   & \rTRUE } \\
strchr-2.c                          & $                 $ & \rAF & \rAF   & \rAF & \rAF   &  2.08 &   0.2 &   \ULTIMATE{  1.6   & \rTRUE } \\

\hline
\end{tabular}
\scriptsize
\[\begin{array}{lll}
\textrm{AlDaFeGo-SAS2010-easy1}: &  -x, -x + 14\cdot z, -x + 18\cdot z, -x + 22\cdot z, -x + 30\cdot z, -x + -2\cdot z, -x + -3\cdot z, -x + -4\cdot z, \\
&  -x + -5\cdot z, -x + -6\cdot z, -x + -17\cdot z, 25 + -x, 29 + -x, 36 + -x, 37 + -x, -x + -38\cdot z   \\
\textrm{ChCoGuSaYa-ESOP2008-easy1}:  &  -x + -z, -x + 3\cdot z, -x + 4\cdot z, -x + -2\cdot z, -x + -3\cdot z, -x + -4\cdot z, -x + -5\cdot z, -x + -6\cdot z,\\
&  -x + -9\cdot z, -x + -22\cdot z, -x + 5\cdot z, -x + 7\cdot z, -x + 15\cdot z, -x + 16\cdot z, 15 + -x, 21 + -x,\\
&  25 + -x, 34 + -x, -x + 36\cdot z \\
\textrm{KrShTsWi-CAV2010-Fig1.c} & 28 + -x, 82 + -x, 88 + -x, 90 + -x, 104 + -x, 118 + -x, 144 + -x, 156 + -x, 212 + -x, 214 + -x,\\
& 228 + -x, 234 + -x, 246 + -x  \\
\textrm{Urban-WST2013-Fig2.c}: & -x1, -x1 + 5\cdot x2, -x1 + 6\cdot x2, -x1 + 7\cdot x2, -x1 + 8\cdot x2, -x1 + 9\cdot x2, -x1 + 10\cdot x2, -x2 
\end{array}\]
\end{center}

\section{Naming abbreviations for \textts{termination-crafted-lit}}
\label{apx:naming}

\begin{center}\scriptsize
\begin{tabular}{ll}
\textts{AlDaFeGo-SAS2010-Fig1.c} & \textts{AliasDarteFeautrierGonnord-SAS2010-Fig1.c} \\
\textts{AlDaFeGo-SAS2010-cousot9.c} & \textts{AliasDarteFeautrierGonnord-SAS2010-cousot9.c} \\
\textts{AlDaFeGo-SAS2010-easy1.c} & \textts{AliasDarteFeautrierGonnord-SAS2010-easy1.c} \\
\textts{AlDaFeGo-SAS2010-easy2-2.c} & \textts{AliasDarteFeautrierGonnord-SAS2010-easy2-2.c} \\
\textts{AlDaFeGo-SAS2010-loops.c} & \textts{AliasDarteFeautrierGonnord-SAS2010-loops.c} \\
\textts{AlDaFeGo-SAS2010-nestedLoop-1.c} & \textts{AliasDarteFeautrierGonnord-SAS2010-nestedLoop-1.c} \\
\textts{AlDaFeGo-SAS2010-random1d-2.c} & \textts{AliasDarteFeautrierGonnord-SAS2010-random1d-2.c} \\
\textts{AlDaFeGo-SAS2010-random2d.c} & \textts{AliasDarteFeautrierGonnord-SAS2010-random2d.c} \\
\textts{AlDaFeGo-SAS2010-speedpldi2.c} & \textts{AliasDarteFeautrierGonnord-SAS2010-speedpldi2.c} \\
\textts{AlDaFeGo-SAS2010-speedpldi3.c} & \textts{AliasDarteFeautrierGonnord-SAS2010-speedpldi3.c} \\
\textts{AlDaFeGo-SAS2010-speedpldi4.c} & \textts{AliasDarteFeautrierGonnord-SAS2010-speedpldi4.c} \\
\textts{AlDaFeGo-SAS2010-wcet2.c} & \textts{AliasDarteFeautrierGonnord-SAS2010-wcet2.c} \\
\textts{AlDaFeGo-SAS2010-while2.c} & \textts{AliasDarteFeautrierGonnord-SAS2010-while2.c} \\
\textts{AlDaFeGo-SAS2010-wise.c} & \textts{AliasDarteFeautrierGonnord-SAS2010-wise.c} \\
\textts{Av-FLOPS2006-Table1.c} & \textts{Avery-FLOPS2006-Table1.c} \\
\textts{BrCoFu-CAV2013-Fig1.c} & \textts{BrockschmidtCookFuhs-CAV2013-Fig1.c} \\
\textts{BrCoFu-CAV2013-Fig9a.c} & \textts{BrockschmidtCookFuhs-CAV2013-Fig9a.c} \\
\textts{BrCoFu-CAV2013-Intro.c} & \textts{BrockschmidtCookFuhs-CAV2013-Introduction.c} \\
\textts{BrMaSi-CAV2005-Fig1-mod.c} & \textts{BradleyMannaSipma-CAV2005-Fig1-modified.c} \\
\textts{BrMaSi-CAV2005-Fig1.c} & \textts{BradleyMannaSipma-CAV2005-Fig1.c} \\
\textts{ChCoFuNiHe-TACAS2014-Intro.c} & \textts{ChenCookFuhsNimkarOHearn-TACAS2014-Introduction.c} \\
\textts{ChCoGuSaYa-ESOP2008-easy1.c} & \textts{ChawdharyCookGulwaniSagivYang-ESOP2008-easy1.c} \\
\textts{ChCoGuSaYa-ESOP2008-easy2.c} & \textts{ChawdharyCookGulwaniSagivYang-ESOP2008-easy2.c} \\
\textts{ChCoGuSaYa-ESOP2008-random1d.c} & \textts{ChawdharyCookGulwaniSagivYang-ESOP2008-random1d.c} \\
\textts{ChCoGuSaYa-ESOP2008-random2d.c} & \textts{ChawdharyCookGulwaniSagivYang-ESOP2008-random2d.c} \\
\textts{ChFlMu-SAS2012-Ex2.20.c} & \textts{ChenFlurMukhopadhyay-SAS2012-Ex2.20.c} \\
\textts{ChFlMu-SAS2012-Ex3.01.c} & \textts{ChenFlurMukhopadhyay-SAS2012-Ex3.01.c} \\
\textts{CoSeZu-TACAS2013-Fig1.c} & \textts{CookSeeZuleger-TACAS2013-Fig1.c} \\
\textts{CoSeZu-TACAS2013-Fig7a.c} & \textts{CookSeeZuleger-TACAS2013-Fig7a.c} \\
\textts{CoSeZu-TACAS2013-Fig7b.c} & \textts{CookSeeZuleger-TACAS2013-Fig7b.c} \\
\textts{CoSeZu-TACAS2013-Fig8a-mod.c} & \textts{CookSeeZuleger-TACAS2013-Fig8a-modified.c} \\
\textts{CoSeZu-TACAS2013-Fig8a.c} & \textts{CookSeeZuleger-TACAS2013-Fig8a.c} \\
\textts{CoSeZu-TACAS2013-Fig8b.c} & \textts{CookSeeZuleger-TACAS2013-Fig8b.c} \\
\textts{GoRe-CAV2006-Fig1a.c} & \textts{GopanReps-CAV2006-Fig1a.c} \\
\textts{HaLaNoRa-SAS2010-Fig1.c} & \textts{HarrisLalNoriRajamani-SAS2010-Fig1.c} \\
\textts{HaLaNoRa-SAS2010-Fig3.c} & \textts{HarrisLalNoriRajamani-SAS2010-Fig3.c} \\
\textts{HeHoLePo-ATVA2013-Fig1.c} & \textts{HeizmannHoenickeLeikePodelski-ATVA2013-Fig1.c} \\
\textts{HeHoLePo-ATVA2013-Fig4.c} & \textts{HeizmannHoenickeLeikePodelski-ATVA2013-Fig4.c} \\
\textts{HeHoLePo-ATVA2013-Fig5.c} & \textts{HeizmannHoenickeLeikePodelski-ATVA2013-Fig5.c} \\
\textts{HeHoLePo-ATVA2013-Fig6.c} & \textts{HeizmannHoenickeLeikePodelski-ATVA2013-Fig6.c} \\
\textts{HeJhMaSu-POPL2002-LockEx.c} & \textts{HenzingerJhalaMajumdarSutre-POPL2002-LockingExample.c} \\
\textts{KrShTsWi-CAV2010-Ex.c} & \textts{KroeningSharyginaTsitovichWintersteiger-CAV2010-Ex.c} \\
\textts{KrShTsWi-CAV2010-Fig1.c} & \textts{KroeningSharyginaTsitovichWintersteiger-CAV2010-Fig1.c} \\
\textts{LaOlRo-CaRu-FMCAD2013-Fig1.c} & \textts{LarrazOliverasRodriguez-CarbonellRubio-FMCAD2013-Fig1.c} \\
\textts{LeHe-TACAS2014-Ex1.c} & \textts{LeikeHeizmann-TACAS2014-Ex1.c} \\
\textts{LeHe-TACAS2014-Ex9.c} & \textts{LeikeHeizmann-TACAS2014-Ex9.c} \\
\textts{LeHe-WST2014-Ex9.c} & \textts{LeikeHeizmann-WST2014-Ex9.c} \\
\textts{LeJoBe-Amram-POPL2001-Ex4.c} & \textts{LeeJonesBen-Amram-POPL2001-Ex4.c} \\
\textts{LeJoBe-Amram-POPL2001-Ex5.c} & \textts{LeeJonesBen-Amram-POPL2001-Ex5.c} \\
\textts{PoRy-LICS2004-Fig1.c} & \textts{PodelskiRybalchenko-LICS2004-Fig1.c} \\
\textts{PoRy-TACAS2011-Fig1.c} & \textts{PodelskiRybalchenko-TACAS2011-Fig1.c} \\
\textts{PoRy-TACAS2011-Fig2.c} & \textts{PodelskiRybalchenko-TACAS2011-Fig2.c} \\
\textts{PoRy-TACAS2011-Fig4.c} & \textts{PodelskiRybalchenko-TACAS2011-Fig4.c} \\
\textts{Ur-WST2013-Fig1.c} & \textts{Urban-WST2013-Fig1.c} \\
\textts{Ur-WST2013-Fig2-mod1000.c} & \textts{Urban-WST2013-Fig2-modified1000.c} \\
\textts{Ur-WST2013-Fig2.c} & \textts{Urban-WST2013-Fig2.c} \\
\textts{UrMi-ESOP2014-Fig3.c} & \textts{UrbanMine-ESOP2014-Fig3.c} \\
\textts{Velroyen.c} & \textts{Velroyen.c} \\
\textts{aviad.c} & \textts{aviad.c} \\
\textts{cstrcmp.c} & \textts{cstrcmp.c} \\
\textts{cstrcspn.c} & \textts{cstrcspn.c} \\
\textts{cstrlen.c} & \textts{cstrlen.c} \\
\textts{cstrpbrk.c} & \textts{cstrpbrk.c} \\
\textts{cstrspn.c} & \textts{cstrspn.c} \\
\textts{genady.c} & \textts{genady.c} \\
\textts{strchr-2.c} & \textts{strchr-2.c} \\

\end{tabular}
\end{center}

\section{Transformation for tracing and truncation}
\label{apx:cfa}

For simplicity, we define this transformation in terms of the widely used notion of control flow automata (CFA):
\begin{definition}[Control-flow automaton~\cite{Henzinger2002}]
A (deterministic) \emph{control flow automaton} denoted $\aut$, is a
tuple $\aut = \langle Q, q_0, X, s, \cfarightarrow{\,} \rangle$ where $Q$ is a
finite set of control locations and $q_0$ is the initial control location, $X$
is a finite set of typed variables, $s$ is the statement language 
and $\cfarightarrow{\,}\subseteq Q \times s \times Q$ is a finite set
of labeled edges.
\end{definition}
The statement language $s$ involves loop/branch-free code, with \texttt{assume} statements.
%\noindent
%We will use the follow statement language:
%\[\begin{array}{lll}
%  % \mid \ttite{b}{e}{e} 
%  e &::=& e \oplus e \mid  c \mid \mathbb{Z} \mid x\\
%  b &::=& b \otimes b \mid \textsf{true} \mid \neg b \\
%  s &::=& s \texttt{ ; } s \mid x := e \mid \ttassume{b} \mid \ttskip
%\end{array}\]
%For lack of space, we omit semantic details of CFAs (see Appendix~\ref{apx:cfasem}). 
We provide the following standard definitions of CFA semantics.
We define a \emph{valuation} of variables 
$\vln : X \rightarrow \mathit{Val}$ to be a mapping from
variable names to values. Let $\Vlns$ be the set of all valuations.
The notation $\vln' \in \sem{s}\vln$ means that executing statement $s$, using
the values given in $\vln$, leads to a new
valuation $\vln'$, mapping variables $X$ to new values. Notations $\sem{e}\vln$ and $\sem{b}\vln$ represent side-effect free numeric and boolean valuations, respectively.
We assume that for every $\vln,s$, that $\sem{s}\vln$ is computed in finite time.
The automaton edges, along with $\sem{s}$, give rise to possible transitions, denoted
$(q,\vln) \cfarightarrow{s} (q',\vln')$ (we omit these rules for lack of space.
A \emph{run} of a CFA is an alternation of automaton states and
valuations: $\run=q_0,\vln_0,q_1,\vln_1,q_2,\dots$
such that $\forall i \geq0.\ (q_i,\vln_i) \cfarightarrow{s} (q_{i+1},\vln_{i+1})$.
%%
%We say that CFA $\aut$ can \emph{reach} automaton state $q$ (\emph{safety}) provided that
%there exists a run $\run= q_0,\vln_0,q_1,\vln_1,...$ such that there is
%some $i \geq 0 $ such that  $q_i=q$.

We now describe the instrumentation and truncation on CFAs.
%and so \emph{state} instrumentation is simply:
%$$\inspect(\sigma) \rightarrow (\pc,x_0,\ldots,x_n) $$
First, we introduce a new statement $\vtracePre{i}$, that is identical to \ttskip, except that it has a side effect output of the tuple
$(pre,i,x_0,\ldots,x_n)$
where $i$ is the loop identifier and $pre$ indicates that the values come from the location right before the loop $i$'s header.
Similarly, we introduce statements $\vtraceBody{i}$ for the location right before the loop $i$'s body and $\vtracePost{i}$
for the location at the loop's exit.

\newcommand\ttctr{\texttt{\_ctr}}
\newcommand\ttbnd{\texttt{\_bnd}}
\begin{definition}[Truncation and tracing] 
We transform the input program CFA $\aut$ to an output CFA $\aut'$. For every loop $i$, having an entry edge from a state $q$ to header node $q_{head}$, we make 
the following transformation.
% where every loop $i$ is transformed as follows:\timos{what is $q$ below? the state right before the the loop header?}
\[\begin{array}{lcl}
	\left\{\begin{array}{l}
	 q \cfarightarrow{s} q_{head} \\
	 q_{head} \cfarightarrow{\ttassume{b}} q_{body}\\
	  \;\;\; q_{body} \cdots \cfarightarrow{} q_{head}\\
	 q_{head} \cfarightarrow{\ttassume{\neg b}} q_{exit} \\
	 \end{array}\right.
&	 \mapsto &
	\left\{\begin{array}{l}
	 q \cfarightarrow{s} q' \cfarightarrow{\ttctr_i = 0\textts{; }\vtracePre{i}} q_{head}, \qquad
	 q_{head} \cfarightarrow{\ttassume{b}} q'' \\
	q'' \cfarightarrow{\ttassume{\ttctr_i = \ttbnd_i}} q_{abort} \\
	 q'' \cfarightarrow{\ttassume{\ttctr_i \neq \ttbnd_i}; \ttctr_i++;} 
   q''' \cfarightarrow{\vtraceBody{i}} 
   q_{body} \cdots \cfarightarrow{} q_{head}\\
	 q_{head} \cfarightarrow{\ttassume{\neg b}} q'''' 
	 \cfarightarrow{\vtracePost{i}} q_{exit} \\
	 \end{array}\right.
\end{array}\]
where primed states $q',q'',$ etc. are fresh states and
$q_{abort}$  corresponds to the exit of \texttt{main}.
\end{definition}
\noindent
Intuitively the transformation consists of two parts: (i) Adding an $\vtracePre{i}$ just before the loop,  $\vtraceBody{i}$ at the beginning of the loop body, and  $\vtracePost{i}$ just outside of the loop. (ii) Introducing a counter $\ttctr_i$ specific to the loop, that is initialized to 0 before the loop, incremented on each iteration of the loop and, if ever the counter reaches the predefined bound $\ttbnd_i$ for the loop, control may break out of the loop (even if the original loop's condition may hold) and immediately exit \texttt{main}. Counter instrumentation is done in other contexts~\cite{gulwani2009speedb}.
% Here we truncate executions.

%
%
%
%\begin{definition}[Traces of instrumented programs]
%\red{todo}.
%different from runs. emit tuples.
%$$(\pc,x_0,\ldots,x_n) \textsf{list}$$
%$$
% \{ (pc,x), (pc,x)... \}
%$$
%\end{definition}
%
%concrete traces? sequences of tuples?

%\section{Precious garbage}
%\input{cfa}
%\section{Appendix}
%
%Text of appendix \ldots

\end{document}